\documentclass{article}
\usepackage{graphicx}  
\usepackage{amsmath}
\usepackage{subcaption}
\usepackage{amssymb}   
\usepackage{bm} 
\usepackage{dcolumn}
\usepackage{color}
\usepackage{mathrsfs}
\usepackage{amsfonts}
\usepackage{varioref}
\usepackage[margin=1.6in]{geometry}
\usepackage{slashed}
\RequirePackage[colorlinks,citecolor=blue,urlcolor=magenta,linkcolor=blue]{hyperref}
\usepackage{comment}
\def\p{\partial}
\def\e{\epsilon}

\def\be{\begin{equation}}
\def\ee{\end{equation}}

\labelformat{section}{Section #1} 
\labelformat{subsection}{Section #1} 
\labelformat{subsubsection}{Section #1}
\labelformat{subsubsubsection}{Section #1}
\labelformat{equation}{Eq.~(#1)} 
\labelformat{figure}{Fig.~#1} 
\labelformat{subfigure}{Fig.~\thefigure#1} 
\labelformat{table}{Tab.~#1} 
\labelformat{appendix}{Appendix #1}

\title{\bf Yukawa coupling, and  inflationary correlation functions for a spectator scalar via stochastic spectral expansion}
\author{$^{1}$Sourav Bhattacharya\footnote{sbhatta.physics@jadavpuruniversity.in}\,\, and $^2$Sudesh Kumar\footnote{sudesh.21phz0007@iitrpr.ac.in}\\
\small{$^1$Relativity and Cosmology Research Centre, Department of Physics, Jadavpur University, Kolkata 700 032, India}\\
\small{$^2$Department of Physics, Indian Institute of Technology Ropar, Rupnagar, Punjab 140 001, India}\\}
\begin{document}
\maketitle
\begin{abstract}
\noindent
We consider a stochastic spectator scalar field coupled to  fermion via the Yukawa interaction, in the inflationary de Sitter background. We consider the fermion to be massless, and  take the  one loop effective potential  found  earlier by using the exact fermion propagator in  de Sitter spacetime. We take the potential for the spectator scalar to be  quintessence-like,  $V(\phi)=\alpha |\phi|^p$ ($\alpha \ensuremath{>} 0,\ p\ensuremath{>} 4$), so that the total effective potential is generically bounded from below for all values of the parameters and couplings, and a late time equilibrium state is allowed. Using next the  stochastic spectral expansion method, we numerically investigate the two point correlation function, as well as the density fluctuations corresponding to the spectator field, with respect to the three parameters of the total effective potential, $\alpha,\ p$ and the Yukawa coupling, $g$. In particular, we find that the power spectrum and the spectral index corresponds to blue tilt with increasing $g$. The three point correlation function and non-Gaussianity corresponding to the density fluctuation has also been investigated.  The increasing Yukawa coupling is shown to flatten the peak of the shape function in the squeezed limit. Also in this limit, the increase in the same is shown to increase the local non-Gaussianity parameter.
\end{abstract}
\vskip .5cm

\newpage

\tableofcontents

\section{Introduction}\label{S1}
The primordial cosmic inflation is a phase of very rapid accelerated expansion of our universe. During this phase the Hubble rate remains nearly constant and the spacetime is well approximated as (quasi)-de Sitter. This inflationary phase solves the puzzles  of standard big bang cosmology, like the horizon problem, the flatness problem  or the rarity of relics like the magnetic monopoles~\cite{Mukhanov:2005sc}.
Inflation can be  achieved by, for example, a slowly rolling scalar field on a potential. Towards the end of inflation, the scalar reaches the minimum of that potential, starts oscillating rapidly leading to the thermalisation of the universe. This leads to the question of cosmic coincidence regarding the current observed value of the dark energy density, see 
e.g.~\cite{Starobinsky:1982ee, Tsamis, Ringeval, Evnin:2018zeo} and references therein.

During this inflationary phase, quantum fluctuations of light scalar fields were created and were stretched to the  deep infrared (IR), super-Hubble scales. It is believed that these fluctuations became classical at large scales through some decoherence mechanism, reentered the horizon after the inflation ends, and eventually grew to build the large scale structures we observe today in the sky. There are various intriguing and open issues regarding the fluctuations of light scalar or gravitational field, including particle creation, non-equilibrium, non-perturbative IR effects in correlation functions and dynamical mass generation at late times. For quantum field  theoretic discussion on these issues, we refer 
our reader to e.g.~\cite{Allen:1985ux, Allen, Floratos, Onemli:2002hr,Weinberg, Brunier:2004sb, Miao:2006pn, Garbrecht:2006jm, Enqvist:2008kt, Onemli:2015pma,  Burgess:2015ajz, Karakaya:2017evp, Kamenshchik:2020yyn, Kamenshchik:2024ybm, Miao:2021gic, Bhattacharya:2022aqi, Bhattacharya:2023yhx, Cabass:2022avo, Brahma:2024yor, Akhmedov:2024npw} and references therein.

It turns out that for a scalar field with canonical kinetic term and a potential which is bounded from below, the late time long wavelength behaviour is stochastic. This corresponds to the existence of a one particle probability distribution function satisfying the Fokker-Planck equation. The stochastic formalism is very efficient to resum the non-perturbative effects in deep IR and hence in correlation functions~\cite{Starobinsky:1986fx, Starobinsky:1994bd}. We further refer our reader to 
~e.g.~\cite{Finelli:2008zg, Martin:2011ib, Motohashi:2012bb, Fujita:2013cna, Fujita:2014tja, Vennin:2015hra, Assadullahi:2016gkk, Firouzjahi:2018vet, Pattison:2019hef, Hardwick:2019uex, Markkanen:2019kpv, Markkanen:2020bfc, Ando:2020fjm, Ebadi:2023xhq, Cruces:2022imf} and references therein for various aspects of stochastic inflation and computations of correlation functions. The various $n$-point correlation functions and associated power spectra are  the most important observables in cosmology.  In particular, the three point correlation function may tell us about departure from standard Gaussian distribution of the fields leading to non-linearity. This has imortant consequence pertaining to the structure formation, see  e.g.~\cite{Gangui:1993tt, Verde:1999ij, Komatsu:2001rj, Maldacena:2002vr, Creminelli:2004yq, Lyth:2005fi, Boubekeur:2005fj, Meerburg:2009ys, Jeong:2009vd, yadav2010primordial, Antoniadis:2011ib, Kehagias:2012pd, Gwyn:2012pb, Kristiano:2023scm}
and references therein for quantum field theoretic as well as stochastic computations. 

In this work we wish to investigate the effect of the Yukawa interaction  on the two and three point correlation functions and associated power spectra for a {\it spectator} scalar field with a generic quintessence-like potential, using the stochastic spectral expansion technique~\cite{Starobinsky:1994bd, Markkanen:2019kpv, Markkanen:2020bfc}. A spectator is not an inflaton and {\it a priori} one expects that its backreaction on the spacetime geometry is ignorable, at least at the initial stages. However, they may leave interesting  physical footprints, e.g.~\cite{Enqvist:2012xn, Friedrich:2019hev, Rigopoulos:2022gso, Glavan:2023lvw, Bhattacharya:2023twz, Wilkins:2023asp} and references therein. We also refer our reader to~ e.g~\cite{Miao:2006pn, Garbrecht:2006jm, Bhattacharya:2023yhx, Bhattacharya:2023twz} for quantum field theoretic discussion on the Yukawa interaction in inflationary de Sitter.\\

\noindent
The rest of the paper is organised as follows. In the next section we discuss the potential for the spectator scalar we will be working in. In particular, the fermionic part of it corresponds to the one loop effective potential derived using the exact propagator in de Sitter in~\cite{Miao:2006pn}. Following \cite{Starobinsky:1994bd, Markkanen:2019kpv, Markkanen:2020bfc}, we describe the stochastic spectral expansion technique in \ref{S3}. We compute the two point correlation, the density correlation, the spectral index in \ref{S4}, and the three point correlation and non-Gaussianity in \ref{S5}. Finally we conclude in \ref{S6} with some future directions. Our analysis will be mostly numerical. To the best of our knowledge, the computation of any three point correlation function using the stochastic spectral expansion method has not been done earlier.  \\

\noindent
We will work with the mostly positive signature of the metric in dimension four, and will set $c=\hbar=1$ throughout.

\section{The fermionic and the total effective potential}
\label{S2}
In this section we shall construct the potential for the stochstic spectator field.
For our purpose, following~\cite{Miao:2006pn} we first wish to briefly review  the derivation of the one loop effective potential for the fermion using its exact propagator in de Sitter. We will work in the inflationary de Sitter background, whose metric  reads
\be 
ds^2= -dt^2 + e^{2Ht}d\vec{x}^2=\frac{1}{H^2\eta^2} \left[-d\eta^2 + d\vec{x}^2 \right]
\label{ds}
\ee
where $H$ is the de Sitter Hubble rate which is a constant. The matter action reads
\begin{eqnarray}
S= \int a^d d^d x \left[-\frac12 (\nabla_{\mu} \varphi)(\nabla^{\mu}\varphi) -\frac{1}{2} \xi_0 R \varphi^2 -\frac12 m^2 \varphi^2 - \frac{\lambda}{4!} \varphi^4 - V(\varphi) + i\bar{\Psi}\slashed{\nabla} \Psi - m_f \bar{\Psi} \Psi -g_0 \bar{\Psi} \Psi \varphi   \right]
\label{cy1}
\end{eqnarray}
where $d=4-\e$, and $V(\varphi)$ is some scalar potential which does not contain any mass or quartic interaction term. We assume that the scalar acts as a background field. While computing the effective action for the fermion, we shall not be concerned about the explicit form of $V(\varphi)$, and will choose it later.  Since we are working with the mostly positive signature of the metric, the anti-commutation relation satisfied by the $\gamma$-matrices read
$$[\gamma^{\mu},\gamma^{\nu}]_+= -\frac{2}{a^{2}} \eta^{\mu\nu}\bf{I}_{d\times d}$$
where $a(\eta)=-1/H\eta$ is the de Sitter scale factor.
Introducing now the field strength renormalisations,
$$\varphi = \sqrt{Z} \phi \qquad \Psi = \sqrt{Z_f} \psi,$$
and making the  decompositions 
$$Z\xi_0 = 0 +\delta \xi, \qquad Z^2 \lambda_0 = \delta \lambda, \qquad Z m^2 =\delta m^2, \qquad  Z\sqrt{Z_f}\, g_0 = g+\delta g, \qquad m_f Z_f = \delta m_f $$
and also writing
$$Z= 1+\delta Z, \qquad Z_f = 1+ \delta Z_f,$$
 \ref{cy1} becomes 
\begin{eqnarray}
S= \int a^d d^d x \left[-\frac{1}{2} (\nabla_{\mu} \phi)(\nabla^{\mu}\phi) -\frac12 \delta \xi R  \phi^2 -\frac{1}{2} \delta m^2 \phi^2 - \frac{\delta \lambda}{4!} \phi^4 - V(\phi) + i\bar{\psi}\slashed{\nabla} \psi + i\delta Z_f \bar{\psi}\slashed{\nabla} \psi  -g\bar{\psi} \psi \phi  
-\delta g\bar{\psi} \psi \phi  \right]
\label{cy3}
\end{eqnarray}
The Feynman  propagator for the fermion field is given by
\begin{eqnarray}
iS[x,x']= a^{-d/2}\left\langle x \left | i\left(i \slashed{\nabla} -g \phi \right)^{-1} \right | x' \right\rangle,
\label{cy4}
\end{eqnarray}
which reads explicitly
\begin{eqnarray}
&&iS[x,x']=  \frac{H^{d-2}}{(4\pi)^{d/2}} \Gamma(d/2-1) \left( ai \slashed{\nabla} \frac{1}{\sqrt{aa'}}  +\sqrt{\frac{a}{a'}} g \phi \ {\bf I}_{d\times d}
 \right)\nonumber\\
&&\times \left[ \left\{ \frac{\Gamma(1-d/2) \Gamma(d/2-1+ ig \phi/H) \Gamma(d/2- i g \phi/H)}{ \Gamma(d/2-1)\Gamma(ig \phi/H) \Gamma(1-ig \phi/H)  }  \, _2F_1\left(\frac{d}{2}-1+ \frac{ig \phi}{H}, \frac{d}{2} -\frac{i g \phi}{H}, \frac{d}{2}, \frac{y}{4} \right)  \right. \right. \nonumber\\&& \left. \left. + \left(\frac{y}{4}\right)^{1-d/2} \,_2F_1 \left(1-\frac{ig\phi}{H}, \frac{i g \phi}{H}, 2-\frac{d}{2}, \frac{y}{4} \right)  \right\} \frac{{\bf I}_{d\times d}
 -\gamma^0}{2}  \right. \nonumber\\&&\left.  + \left\{ \frac{\Gamma(1-d/2) \Gamma(d/2-1- ig \phi/H) \Gamma(d/2+ i g \phi/H)}{ \Gamma(d/2-1)\Gamma(-ig \phi/H) \Gamma(1+ig \phi/H)  }  \, _2F_1\left(\frac{d}{2}-1- \frac{ig \phi}{H}, \frac{d}{2} +\frac{i g \phi}{H}, \frac{d}{2}, \frac{y}{4} \right)  \right. \right. \nonumber\\&& \left. \left. + \left(\frac{y}{4}\right)^{1-d/2} \,_2F_1 \left(1+\frac{ig\phi}{H}, -\frac{i g \phi}{H}, 2-\frac{d}{2}, \frac{y}{4} \right)  \right\} \frac{{\bf I}_{d\times d}
 +\gamma^0}{2}   \right]
\label{cy5}
\end{eqnarray}
The de Sitter invariant distance function appropriate for the Feynman propagator is given by
$$y(x,x')= aa' \left[ (|\eta -\eta'|-i\e)^2 + |\vec{x}-\vec{x'}|^2 \right]$$
\ref{cy5} in the coincidence limit reads
\begin{eqnarray}
&&iS[x,x]=  \frac{ g \phi H^{d-2}}{(4\pi)^{d/2}} \frac{\Gamma(d/2+ i g \phi/H)  \Gamma(d/2- i g \phi/H) \Gamma(1-d/2)}{\Gamma(1+ i g\phi/H) \Gamma(1- ig \phi/H)} \times {\bf I}_{d\times d}
\label{cy6}
\end{eqnarray}

The one loop effective action is obtained by tracing out the fermion degrees of freedom. As we have stated earlier, the scalar is assumed to be mostly a background field, and at the leading order we ignore its quantum fluctuations. This leads to the equation of motion for the scalar
\begin{eqnarray}
\frac{1}{a^d}\p_{\mu} \left(a^{d-2}\eta^{\mu\nu}\p_{\nu} \phi\right) - \left((12 H^2 \delta \xi+\delta m^2) \phi + \frac{\delta \lambda}{3!} \phi^3 +  V'(\phi)\right)  + g {\rm Tr} \left [iS[g \phi](x,x) \right ] =0  
\label{cy8}
\end{eqnarray}
which gives the derivative of the total effective potential 
\begin{eqnarray}
V'_{\rm eff}(\phi)=   V'(\phi) + \left(12 H^2 \delta \xi + \delta m^2\right)\phi + \frac{\delta \lambda}{3!}\phi^3- g {\rm Tr} \left [iS[g \phi](x,x) \right] 
\label{cy9}
\end{eqnarray}
The renormalised effective potential for fermion is found by integrating the last term on the right hand side and then by choosing the counterterms appropriately in~\ref{cy3}, so that~\cite{Miao:2006pn} (see also~\cite{Candelas})
\begin{eqnarray}
V^{\rm eff}_{\rm fermion}(\phi)= -\frac{H^4}{8\pi^2}\left[ 2\gamma \left(\frac{g\phi}{H} \right)^2 - \left(\zeta(3) -\gamma\right)\left(\frac{g\phi}{H} \right)^4 + 2 \int_0^{g\phi/H} dx\, x(1+x^2) \left( \psi_p(1+ix)+ \psi_p (1-ix) \right) \right]   
\label{cy10}
\end{eqnarray}
where $\gamma$ is the Euler constant and $\psi_p$ is the digamma function,
$$\psi_p(1+x)= \frac{d}{dx} \ln \Gamma(1+x)$$
To keep the picture minimalistic, the {\it finite} quadratic and quartic terms of \ref{cy10} can  further be absorbed by appropriate choices of finite counterterms $\delta m^2$ and $\delta \lambda$,
\begin{eqnarray}
\delta m^2 = \frac{\gamma g^2 H^2}{2\pi^2} \qquad \delta \lambda = -\frac{3g^4\left(\zeta(3) -\gamma\right) }{\pi^2}
\label{cy10'}
\end{eqnarray}
 This leads to the total effective potential  
\begin{eqnarray}
&&V_{\rm eff}(\phi)=  V(\phi ) -\frac{H^4}{4\pi^2}  \int_0^{g\phi/H} dx\, x(1+x^2) \left( \psi_p(1+ix)+ \psi_p (1-ix) \right),
\label{cy11}
\end{eqnarray}
\begin{figure}[ht]
    \centering
    \includegraphics[width=0.45\textwidth]{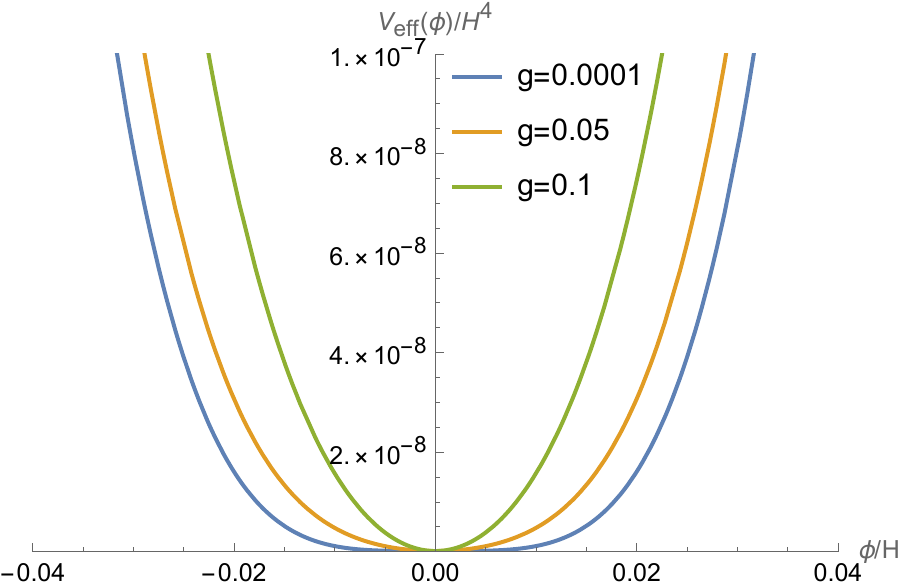} %
    \caption{\small \it Plot of the effective potential \ref{cy11}, \ref{scpot}, with parameters $\bar{\alpha}=0.1$, $p=4.001$ and different values of the Yukawa coupling.}
    \label{fig:effectove-potential}
\end{figure}
 The scalar field's evolution is governed by the derivative of the above effective potential $V_{\rm eff}(\phi)$,
\begin{eqnarray}
\frac{1}{a^d}\p_{\mu} \left(a^{d-2}\eta^{\mu\nu}\p_{\nu} \phi\right) - V'(\phi ) +\frac{g^2 H^2 \phi}{4\pi^2}  \left(1+\left(\frac{g\phi}{H}\right)^2\right) \left[ \psi_p\left(1+ \frac{ig\phi}{H}\right)+ \psi_p \left(1- \frac{ig\phi}{H}\right) \right]=0  
\label{cy12}
\end{eqnarray}
 In order to go beyond the de Sitter, one can replace the de Sitter Hubble rate $H$ by that of the quasi-de Sitter, $H(t)= \sqrt{(\Lambda+8\pi G V(\phi))/3}$ in \ref{cy10}. This approximation will be  valid as long as the slow roll or ultra-slow roll phase prevails, where the variation of $H$ is small. However, we shall focus on a strictly constant  $H$  in this work. The scalar field dynamics from \ref{cy12} for $V(\phi)=0$ can be seen in~\cite{Miao:2006pn}.    \\

\noindent
We now need to make a specific choice of the scalar potential $V(\phi)$. We note here the usual   problem with the Yukawa part of  in \ref{cy11}.  By using the Stirling formula for the digamma function,
\begin{eqnarray}
\psi_p(1+x)= \ln x +\frac{1}{2x}- \sum_{k=1}^{\infty}\frac{B_{2k}}{2k x^{2k}}
\end{eqnarray}
where $B_{2k}$'s are Bernoulli's numbers, we see that for large values of the scalar field ($\phi/H \gg 1$), the leading behaviour of the fermionic part of the effective potential is given by $\sim - g^2\phi^4 \ln |\phi/H|$. Thus it is unbounded from below and no late time equilibrium state for the scalar field is possible for $V^{\rm eff}_{\rm fermion}$ alone. One way to tackle this issue would be to take the scalar field to be quantum with $V(\phi)= \lambda \phi^4$ ($\lambda\ensuremath{>}0$).  Then the effective potential for it for large field values would behave as $\sim +\lambda\phi^4 \ln |\phi/H|$.  Thus if the Yukawa coupling is `sufficiently small', one can have a total effective potential which is bounded from below. However,  we have assumed the scalar to be a background field here, and have ignored its quantum fluctuations at the leading order.  This allows us to choose the $V(\phi)$ term in \ref{cy11} as a  {\it generic}, quintessence-like potential, 
\be
V(\phi) = \alpha |\phi|^p \qquad (\alpha \ensuremath{>}0, \,p\ensuremath{>}4), 
\label{scpot}
\ee 
so that the {\it total} effective potential \ref{cy11} is bounded from below for all values of $\phi$, or various parameters and couplings. The variation of $V_{\rm eff}(\phi)$ has been depicted in~\ref{fig:effectove-potential}. With this choice, we wish to compute various scalar correlation functions using the stochastic spectral expansion formalism.

Finally, we also note that the standard Coleman-Weinberg potential for the Yukawa theory depends upon the arbitrary renormalisation scale $\mu$ as $\sim\phi^4 \ln \phi/\mu$. Thus for $\mu$ extremely large or small, the potential becomes ill-defined, apparently destroying the perturbation theory. A  consistent remedy to this problem is the renormalisation group (RG) improvement of the effective potential, e.g.~\cite{{Hardwick:2019uex}}. Thus perhaps we may speculate that adding the $\alpha |\phi|^p$ term to our one loop effective potential has some  analogy with that of the RG improvement. The obvious and best way to verify this would be to come up with a model whose effective potential or the RG improved version of it  generates a $|\phi|^p$ term along with the standard Yukawa part. While an  explicit such model remains elusive to us so far, we hope attempting with some non-canonical derivative term, or non-minimal coupling could be worth trying. However, one needs to be careful about the renormalisability and hence a correct evaluation of the $\beta$-functions as well. We reserve this issue for a future work.

\section{The  stochastic formalism}
\label{S3}
Let us now review very briefly  the basics of the   stochastic formalism, originally developed in~\cite{Starobinsky:1986fx, Starobinsky:1994bd} (see also \cite{Cruces:2022imf} for a review).
 At late times, the scalar field's mode is divided into two parts -- one is the stretched, long wavelength super-Hubble part and another one is the small, short-wavelength ultraviolet fluctuation. It is the first part whose dynamics is most important for inflationary cosmology. This part, say $\bar{\phi}(t,\vec{x})$, satisfies the Langevin equation under slow roll approximation, 
\begin{equation}\label{langevin equation}
    \dot{\bar{\phi}}+\frac{V'(\bar{\phi})}{3H}=\Sigma(t,\vec{x})
\end{equation}
 $\bar{\phi}(t,\vec{x})$ is coarse grained over a volume larger that what is enclosed  by the Hubble horizon, and hence we have ignored its spatial variation. $\Sigma(t,\vec{x})$  provides the short wavelength random quantum fluctuations, or the white noise,
\begin{equation}\label{delta correlated noise}
    \left\langle \Sigma(t_{1},\vec{x})\Sigma(t_{2},\vec{x})  \right\rangle=\frac{H^{3}}{4\pi^{2}}\delta(t_{1}-t_{2})
\end{equation}
Thus in \ref{langevin equation}, the derivative of the potential provides the drift, whereas the white noise term provides the random kicks. The field $\bar{\phi}$ is classical but random (stochastic), and a one particle probability distribution function can be associated with it. From now on and without any loss of generality, we will write $\bar{\phi}(x,t)=\phi$. Then from \ref{langevin equation}, \ref{delta correlated noise}, the one particle probability distribution function, say $\rho(\phi,t)$, is seen to satisfy the  Fokker-Planck equation~\cite{Starobinsky:1994bd}
\begin{equation}\label{fokker Planck equation}
    \frac{\partial \rho(\phi,t)}{\partial t}=\frac{1}{3H}\frac{\partial}{\partial \phi}(V'(\phi)\rho(\phi,t))+\frac{H^{3}}{8\pi^{2}}\frac{\partial^{2}\rho(\phi,t)}{\partial \phi^{2}}
\end{equation}
 \ref{fokker Planck equation} can be solved by a general solution known as the spectral expansion~\cite{Starobinsky:1994bd},
\begin{equation}\label{solution of fpe}
    \rho(\phi,t)=e^{-\frac{4\pi^{2}V(\phi)}{3H^{4}}}\sum_{n}a_{n}\psi_{n}(\phi)e^{-E_{n}t}
\end{equation}
where $a_{n}$'s are the expansion coefficients, and $\psi_{n}(\phi)$, $E_{n}$ are the eigenfunctions and eigenvalues  satisfying the effective Schr\"odinger-like  equation
\begin{equation}\label{eq:schrodinger equation}
    \biggl[-\frac{1}{2}\frac{\partial^{2}}{\partial \phi^{2}}+W(\phi)\biggl]\psi_{n}(\phi)=\frac{4\pi^{2}E_{n}}{H^{3}}\psi_{n}(\phi)
\end{equation}
where we have abbreviated
\begin{equation}\label{effective potential}
    W(\phi)=\frac{1}{2}\big[\nu'^2(\phi)-\nu''(\phi)\big], \qquad \nu(\phi)\equiv \frac{4\pi^{2}}{3H^{4}}V(\phi)
\end{equation}
The eigenfunctions are taken to be orthonormal and complete
\begin{equation}\label{orthonormal condition}
\int d\phi \psi_{m}(\phi)\psi_{m'}(\phi)=\delta_ {m m'} \qquad \sum_{m}\psi_{m}(\phi)\psi_{m}(\phi')=\delta(\phi-\phi')
\end{equation}
For $E_{0}=0 $, there exists a  stationary solution
\begin{equation}\label{stationary solution}
    \psi_{0}(\phi)\propto \exp\biggl(-\frac{4\pi^{2}}{3H^{4}}V(\phi)\biggr)
\end{equation}
corresponding to the well known equilibrium one-point probability distribution function at late times
\begin{equation}\label{one point pdf}
  \rho_{\rm eq}(\phi)=N \exp\biggl[-\frac{8\pi^{2}}{3H^{4}}V(\phi)\biggr]
\end{equation}
where $N$ is the normalisation constant. 

Let us recall that while deriving the effective potential in the preceding section, we ignored the short wavelength quantum fluctuations mentioned above at the leading order, corresponding to the fact that they  are small. In the next to the leading order, one needs to consider the interactions between the fermion, the long wavelength part of the scalar and those short wavelength modes. However, we shall not consider such  interactions in this work. 

With these, we are now ready  to go into the computation of the two and three point correlation function for \ref{cy11}, \ref{scpot}, using the stochastic spectral expansion method. \\

\noindent
The stoachastic spectral expansion method is a powerful tool for calculating correlation functions, see e.g.~\cite{Markkanen:2019kpv, Markkanen:2020bfc} and references therein for detail. Let us  consider the two-point correlation function for any arbitrary function $f(\phi)$ of the scalar field, 
\begin{equation}\label{two point function}
    \xi^{(2)}_f(t_{1},t_{2};\vec{x}_{1},\vec{x}_{2})\equiv  \langle f( \phi(t_{1},\vec{x}_{1}))f( \phi(t_{2},\vec{x}_{2}))\rangle
\end{equation}
A particular interest will be the temporal coincidence limit ($t_1=t_2=t$, say),  
\be
\xi_f^{(2)}(t, |\vec{x}_1-\vec{x}_2|)
=\langle f(\phi(t,\vec{x}_1)) f(\phi(t,\vec{x}_2))\rangle
\label{autocorrletion}
\ee
where the spatial argument of the left hand side follows from the spatial homogeneity and isotropy of the background cosmological spacetime.

Probability distributions between two time slices are related to each other by  
\begin{equation}\label{linearity}
\tilde{\rho}(t+\Delta t;\phi)=\int d\phi_0 \tilde{U}(\Delta t;\phi,\phi_0)\tilde{\rho}(t;\phi_0),
\end{equation}
where $\tilde{U}(\Delta t;\phi,\phi_0)$ is the kernel  (or its matrix elements) for the time evolution of the spectator. While the kernel is written with respect to the coordinate basis in standard quantum mechanics, we have represented it here using  the field basis, appropriate to our present scenario. The kernel also satisfies the Fokker-Planck equation,~\ref{fokker Planck equation}, where  $ \tilde{\rho}(t;\phi)$ is defined as
\begin{equation}\label{eq:onepointpdf}
\tilde{\rho}(t;\phi)=e^{\nu(\phi)}\rho(t;\phi)
\end{equation}
where $\nu(\phi)$ is given by \ref{effective potential}.
Using the above, \ref{linearity} can be rewritten as 
\begin{eqnarray}
\rho(t+\Delta t;\phi)
&=&
\int d\phi_0 U(\Delta t;\phi,\phi_0)
\rho(t;\phi_0)
\end{eqnarray}
where we have written
\begin{equation}
\label{eq:transfermatrix}
U(\Delta t;\phi,\phi_0)\equiv
e^{-\nu(\phi)}\tilde{U}(\Delta t;\phi,\phi_0)
e^{\nu(\phi_0)}
\end{equation}
The  kernel satisfies the initial conditions
\begin{eqnarray}\label{eq:intitialofu'}
    U(0;\phi,\phi_0)=\tilde{U}(0;\phi,\phi_0)=\delta(\phi-\phi_0)
  \end{eqnarray}
Using the above equation and  \ref{orthonormal condition}, one can write
\begin{eqnarray}\label{eq:intitialofu}
     \hspace{1cm} \tilde{U}(t;\phi,\phi_0)=\sum_n e^{-E_nt}\psi_n(\phi)\psi_n(\phi_0)
\end{eqnarray}
The two-point correlation function can be written in terms of the  kernel and  the one-point equilibrium probability distribution function~\ref{one point pdf}, as
\begin{eqnarray}
{\xi_f^{(2)}(t)}&=&
\int d\phi d\phi_0 \ \psi_0(\phi_0)f(\phi_0) 
\tilde{U}(t;\phi,\phi_0)\psi_0(\phi) f(\phi),
\end{eqnarray}
which equals
\begin{equation}
\xi_f^{(2)}(t)=\sum_n f_n^2 e^{-E_nt},
\label{eq:spectral}
\end{equation}
where the $f_{n}$'s are called the {\it spectral coefficients}.  In terms of the eigenfunctions of the Schr\"{o}dinger-like eigenvalue equation \ref{eq:schrodinger equation}, one has
\begin{equation}
f_n\equiv\int d\phi \psi_0f(\phi)\psi_n\,.\label{coefficient}
\end{equation}
Using these, the two point correlation function of~\ref{autocorrletion} becomes~\cite{Starobinsky:1994bd},
\be
\xi^{(2)}_f(\vec{x}_1,\vec{x}_2,t)=
\sum_n \frac{f_n^2}{\left(R H
\right)^{2E_{n}/H}} 
\label{equ:timetospace}
\ee
where  $R=a(t)|\vec{x}_{1}-\vec{x}_{2}|$ is the proper distance at cosmological time $t$ and $E_n$'s are the eigenvalues of the effective Schr\"{o}dinger equation~\ref{eq:schrodinger equation}. The power spectrum of the correlation function is defined as 
\be
\mathcal{P}^{(2)}_f(k)=\frac{k^3}{2\pi^2}\int d^3\vec{x} \ e^{i\vec{k}\cdot(\vec{x}_1-\vec{x}_2)}\xi_f^{(2)}(|\vec{x}_1-\vec{x}_2|,t)
\ee
where $k=|\vec{k}|$. The above takes the form
\be
\mathcal{P}^{(2)}_f(k)=\frac{2}{\pi}\sum_n f_n^2 \Gamma\bigg(2-\frac{2E_n}{H}\bigg)\sin\frac{E_n\pi}{H}\bigg(\frac{k}{H a}\bigg)^{2E_n/H}
\label{spectrumfull}
\ee
In the late time, deep infrared limit we are interested in, we have $k/a\ll H $, and  the power spectrum can be shown to behave  at the leading order
\begin{equation}
\begin{split}
{\mathcal P}_f^{(2)}(k)\big\vert_{\rm leading}
\simeq 
f_n^2\left(n_s-1\right) \left(\frac{k}{H a}\right)^{n_s-1} 
\end{split}
\label{eq:spectrumP}
\end{equation}
where 
\be
n_s-1 = \frac{2E_n}{H}
\label{ns}
\ee
is called the spectral index. The $E_n$ and $f_n$ appearing above stands for the leading behaviour in \ref{eq:spectrumP}. For $n_s=1$, the power spectrum is exactly momentum  scale invariant. The blue tilted spectrum corresponds to $n_s\ensuremath{>}1$, whereas the spectrum with $n_s\ensuremath{<}1$ is called red tilted. Thus our task below will be to compute the eigenvalues $E_n$ and the spectral coefficients $f_n$. Since we will be working in the deep infrared or the super-Hubble limit, we need to compute  those values appropriate for  the leading behaviour.

\section{The eigenvalue equation and the two point correlation function}\label{S4}

\begin{figure}[htp]
    \centering
\includegraphics[width=7.6cm]{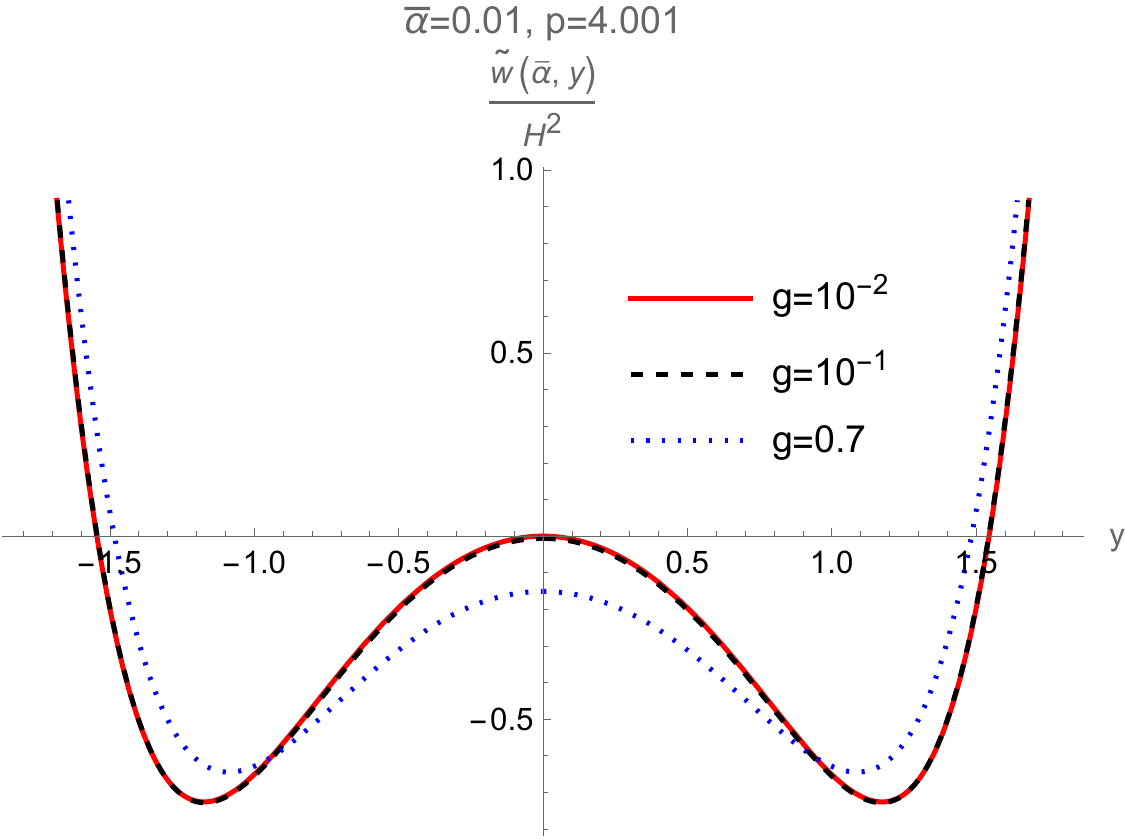}
\caption{\small \it The variation of the quantity $\bar{W}(\bar{\alpha},y)$, \ref{wphi2}, which appears in the effective Schr\"odinger equation \ref{eq:schrodinger equation}. The double well behaviour comes from the symmetry of our effective potential, \ref{eq:eff-pot}.}
\label{fig:weff}
\end{figure}

\begin{figure}[htbp]
    \centering
    \begin{subfigure}[b]{0.328\textwidth}
        \includegraphics[width=\textwidth]{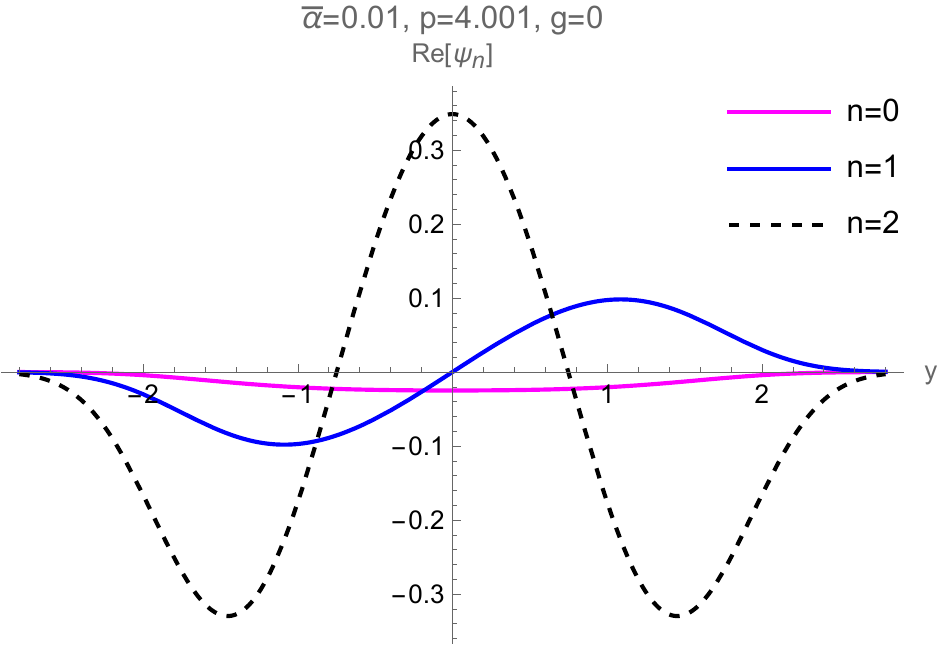}
        \caption{}
        \label{}
    \end{subfigure}
    \hfill
    \begin{subfigure}[b]{0.328\textwidth}
        \includegraphics[width=\textwidth]{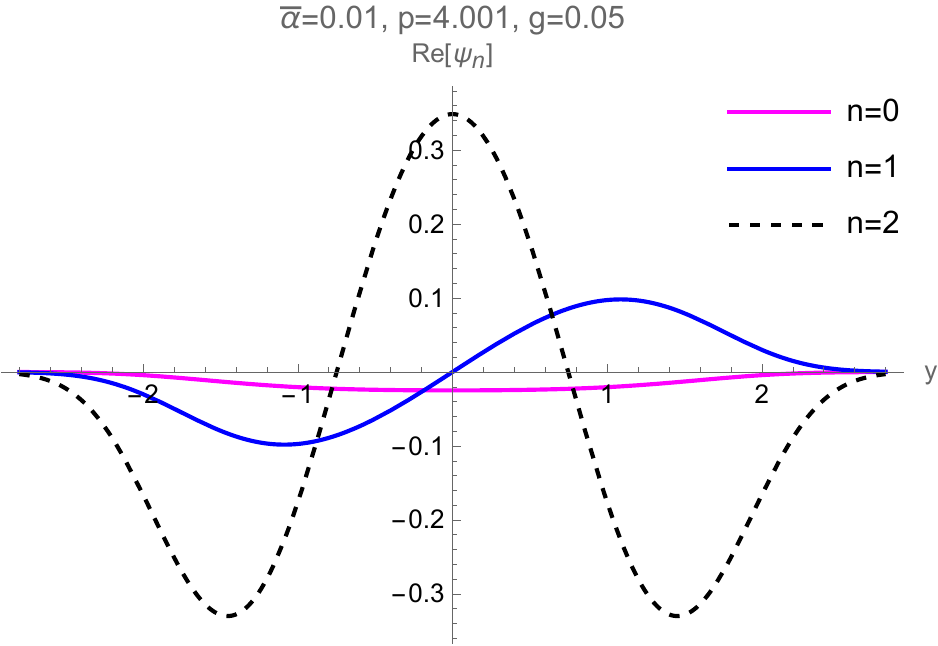}
        \caption{}
        \label{}
    \end{subfigure}
    \hfill
    \begin{subfigure}[b]{0.328\textwidth}
        \includegraphics[width=\textwidth]{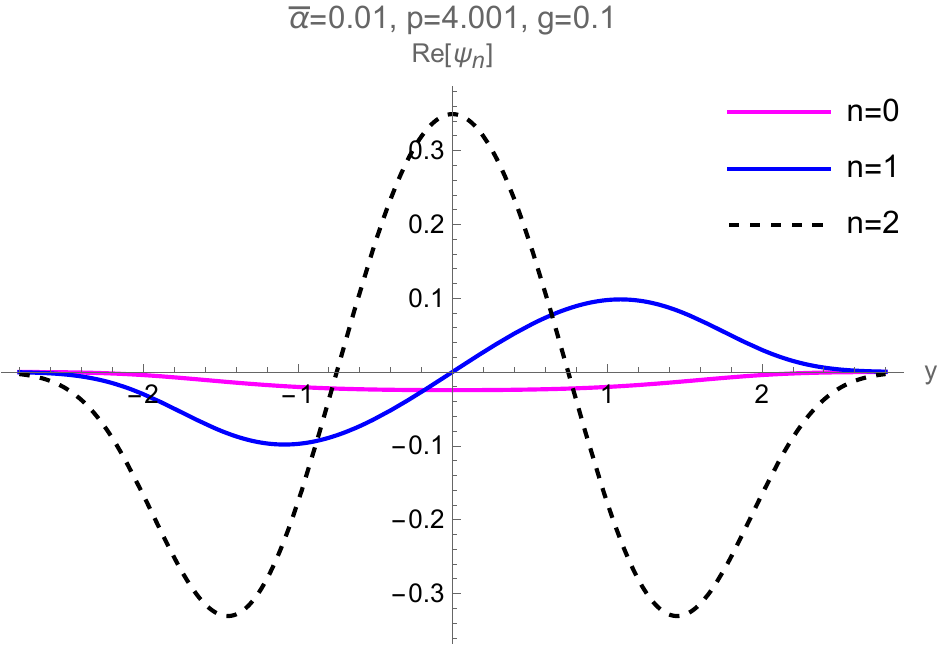}
        \caption{}
        \label{}
    \end{subfigure}
     \caption{\small \it Plots of the lowest three  of the wavefunctions $\psi_n(\bar{\alpha},p,g,y)$,\ref{eq:schrodinger equation}, \ref{eq:eff-pot}, \ref{wphi2}. Parameters $\bar{\alpha}$ and $p$ are held fixed and the Yukawa coupling $g$ is varied. The wavefunction is complex, and we have plotted its real part. The imginary part shows the same qualitative behaviour.}
    \label{fig:plot14}
\end{figure}

 We now wish to find out the eigenvalues $E_n$ appearing in the effective Schr\"{o}dinger 
equation~\ref{eq:schrodinger equation}, with
our effective potential
\begin{eqnarray}
\label{eq:eff-pot}
&&V_{\rm eff}(\phi)=  \alpha\lvert \phi \rvert ^{p} -\frac{H^4}{4\pi^2}  \int_0^{g\phi/H} dx\, x(1+x^2) \left( \psi_p(1+ix)+ \psi_p (1-ix) \right)
\end{eqnarray}
By defining the dimensionless variable and parameter
\begin{eqnarray*}
    y=\frac{\phi}{H}, & \bar{\alpha}=\frac{\alpha}{H^{4-p}},
\end{eqnarray*}
we compute for \ref{effective potential}
\begin{eqnarray}\label{eq:nud}
  &&  \nu^{\prime}(\bar{ \alpha},y)=\frac{4\pi^2}{3}
\left[\bar{ \alpha} p|y|^{p-1}{\rm sgn}(y)-\frac{g^2 y}{4\pi^2}\left(1+g^2 y^2\right)\left( \psi_{p}\left(1+igy\right)+\psi_{p}\left(1-igy\right)\right) 
\right]\nonumber\\
   &&     \nu^{\prime\prime}(\bar{ \alpha},y)=\frac{4\pi^2}{3}
\left[\bar{\alpha} p(p-1)|y|^{p-2}- \frac{g^4y^2}{2\pi^2}\left(\psi_{p}\left(1+igy\right)+\psi_{p}\left(1-igy\right)\right) \right.\nonumber\\
&&\left.-\frac{g^2}{4\pi^2}\left(1+g^2y^2\right)\left(\psi_{p}\left(1+igy\right)+\psi_{p}\left(1-igy\right)\right)-\frac{i g^3 y}{4\pi^2}\left(1+g^2y^2\right)\left(\psi_{p}^{(1)}\left(1+igy\right)-\psi_{p}^{(1)}\left(1-igy\right)
\right)\right]
\end{eqnarray}
Thus we have in \ref{eq:schrodinger equation},
\begin{eqnarray}\label{wphi2}
   & \bar{W}(\bar{\alpha},y) = \frac{H^{2}}{2}\Biggl[ \left(\frac{4\pi^2}{3}\right)^2 \left\{ \bar{\alpha}^2 p^2 | y|^{2(p-1)} + \frac{g^4 y^2}{4\pi^2} \left(1 +g^2y^2\right)^{2}\left[ \psi_{p}\left(1+{igy}\right) + \psi_{p}\left(1-{igy}\right) \right)^{2}\right. \nonumber\\
    &\left. - \frac{g^2\bar{\alpha}p y}{2\pi} | y|^{p-1}{\rm sgn}(y) \left[ \psi_{p}\left(1+{igy}\right) + \psi_{p}\left(1-{igy}\right) \right] \right\}
     - \frac{4\pi^2 }{3}\left\{ \bar{\alpha }p(p-1) |y|^{p-2} - \frac{g^2 y^{2}}{2\pi} \left[ \psi_{p}\left(1+{igy}\right) + \psi_{p}\left(1-{igy}\right) \right]\right. \nonumber\\
    &\left.- \frac{g^2}{4\pi^2} \left(1+g^2y^2\right) \left[ \psi_{p}\left(1+{igy}\right) + \psi_{p}\left(1-{igy}\right) \right] - \frac{ig^3y}{4\pi^2}\left(1+g^2y^2\right)\left( \psi_{p}^{(1)}\left(1+{igy}\right) - \psi_{p}^{(1)}\left(1-{igy}\right) \right) \right\} \Biggr]
\end{eqnarray}
We have plotted the variation of $\bar{W}(\bar{\alpha},y)$ with respect to the dimensionless quantity $y$ in \ref{fig:weff}, showing a  double well feature, owing to the fact that $V_{\rm eff}(\phi)$ is even in $\phi$. Thus the corresponding eigenfunctions $\psi_n(\phi)$, \ref{eq:schrodinger equation}, will be either even or odd. The wavefunction turns out to be complex and we have plotted its real part in \ref{fig:plot14}. The imaginary part shows the same qualitative behaviour.
The first few eigenvalues $E_n$ for \ref{wphi2} put into \ref{eq:schrodinger equation} has been found numerically using MATHEMATICA and  depicted generically in \ref{tab:example}, for some parameter values.   \\
\begin{table}[ht]
    \centering
    \resizebox{\textwidth}{!}{%
    \begin{tabular}{|c|c|c|c|c|}
        \hline
        \rule{0pt}{18pt} $E_n/H$ &{ $\bar{\alpha}=0.01$, $p=4.001$, $g=0$} & $\bar{\alpha}=0.01$, $p=4.001$, $g=0.01$ & $\bar{\alpha}=0.01$, $p=4.001$, $g=0.1$ & $\bar{\alpha}=0.01$, $p=4.001$, $g=0.7$\\ \hline
        $E_0/H$ &{0} &  0 & 0 & 0 \\ \hline
        $E_1/H$ &{0.020831}& 0.020831& 0.020827 & 0.025459 \\ \hline
        $E_2/H$ &{0.055020}&0.055024 & 0.055047 & 0.059024 \\ \hline
        $E_3/H$ &{0.097757}&0.097757 & 0.097819 & 0.099668 \\ \hline
        $E_4/H$ &{0.147879}&0.147880 & 0.147976 & 0.146121 \\ \hline
        $E_5/H$ &{0.204475}&0.204477 & 0.204607 & 0.197601 \\ \hline
    \end{tabular}%
    }
    \caption{\small \it First few eigenvalues after solving the effective Schr\"{o}dinger equation~\ref{eq:schrodinger equation}, \ref{wphi2}, for some values of the relevant parameters.  }
    \label{tab:example}
\end{table}

\begin{figure}[hbt!]
    \centering
    \begin{subfigure}[b]{0.40\textwidth}
        \centering
        \includegraphics[width=\textwidth]{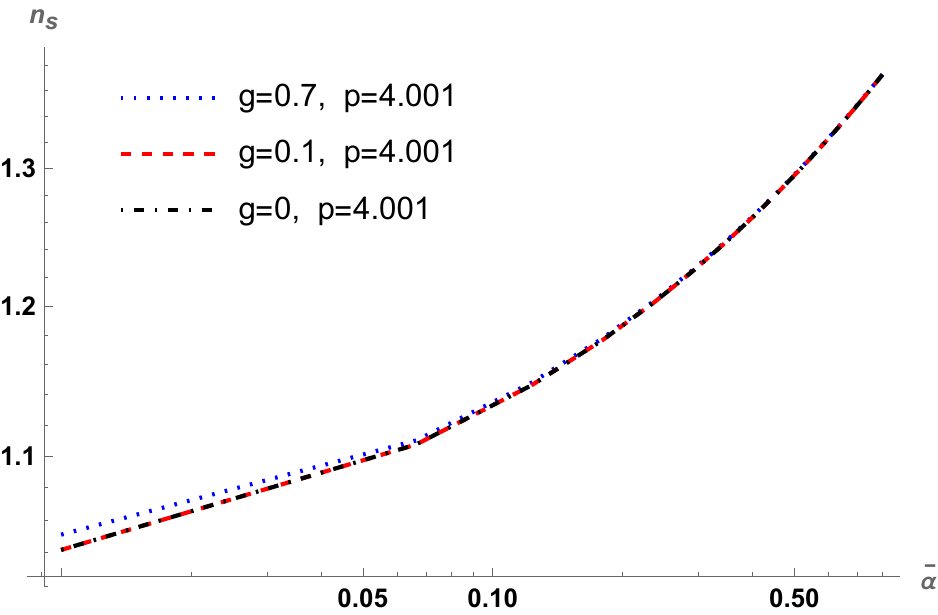}
    \end{subfigure}
    \hfill
    \begin{subfigure}[b]{0.40\textwidth}
        \centering
        \includegraphics[width=\textwidth]{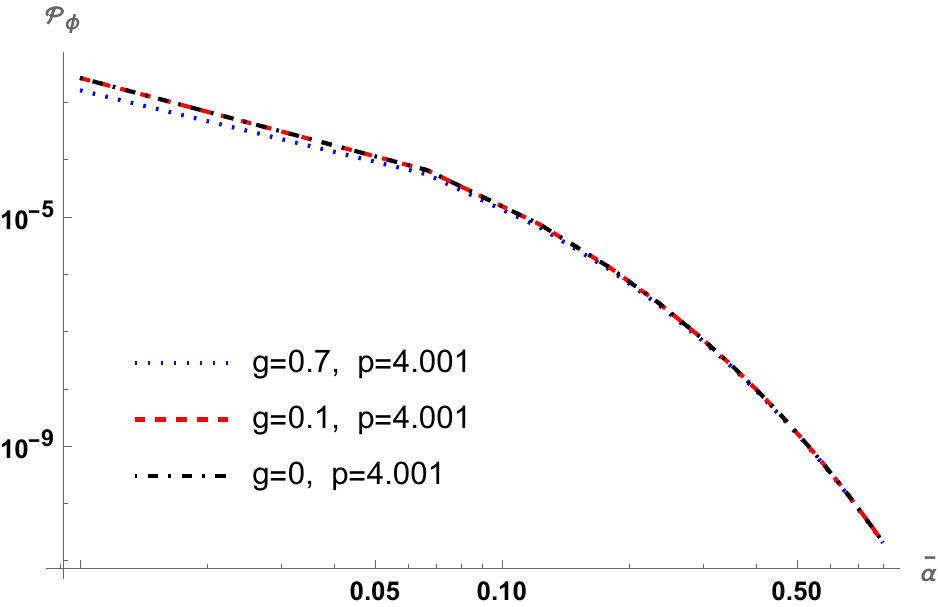}
    \end{subfigure}
    \caption{\small \it Variation of the spectral index and the power spectrum, respectively  \ref{spec}, \ref{P1}, with respect to the dimensionless parameter $\bar\alpha$, \ref{wphi2}. We have taken different values of the Yukawa coupling but the parameter $p$ is held fixed.   }
    \label{fig:both1}
\end{figure}

\begin{figure}[hbt!]
    \centering
    \begin{subfigure}[b]{0.45\textwidth}
        \centering
        \includegraphics[width=\textwidth]{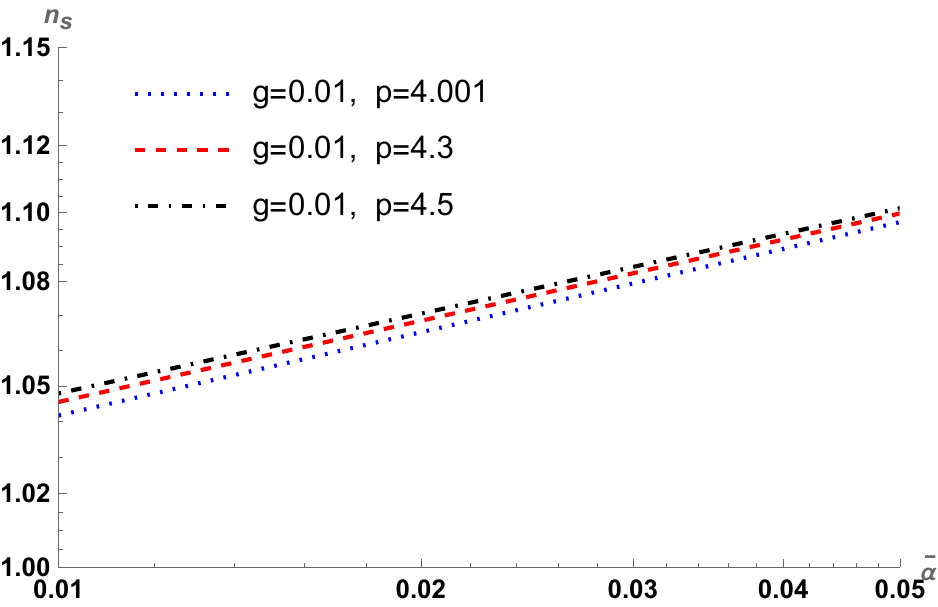}
    \end{subfigure}
    \hfill
    \begin{subfigure}[b]{0.45\textwidth}
        \centering
        \includegraphics[width=\textwidth]{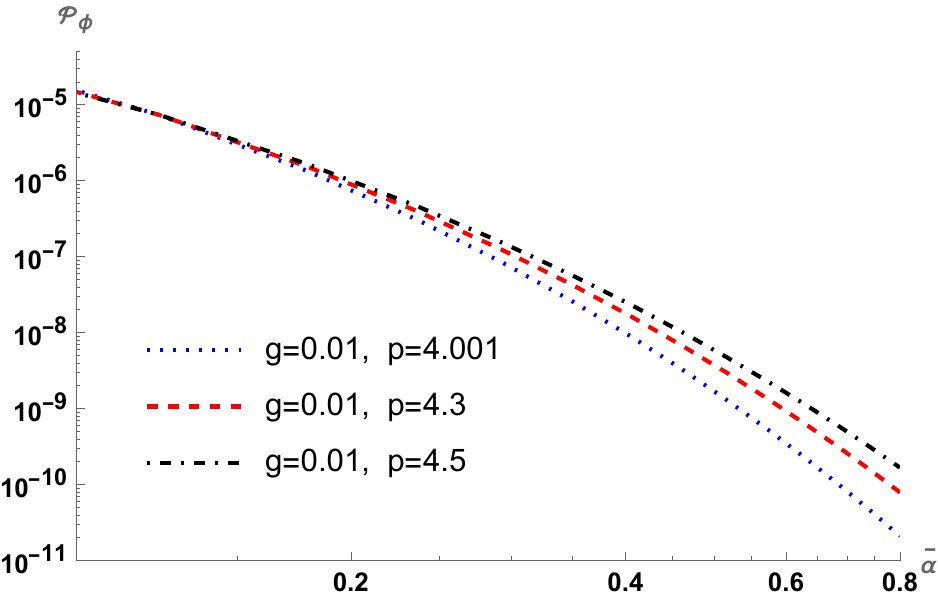}
    \end{subfigure}
    \caption{\small \it Variation of the spectral index and the power spectrum, respectively  \ref{spec}, \ref{P1}, with respect to the dimensionless parameter $\bar\alpha$, \ref{wphi2}. We have taken different values of the parameter $p$, whereas the Yukawa is held fixed.}
    \label{fig:both4}
\end{figure}

\begin{figure}[hbt!]
    \centering
    \begin{subfigure}[b]{0.45\textwidth}
        \centering
        \includegraphics[width=\textwidth]{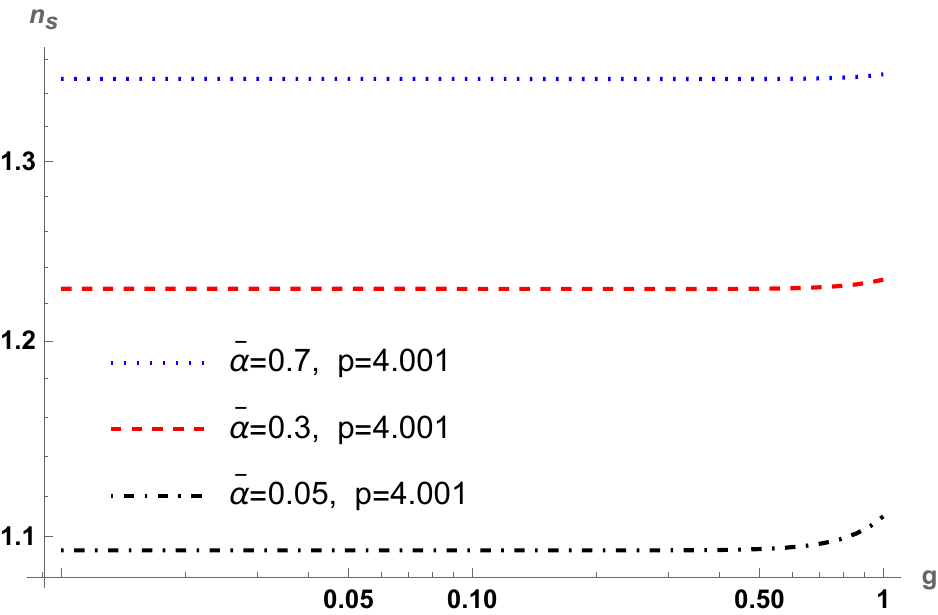}
    \end{subfigure}
    \hfill
    \begin{subfigure}[b]{0.45\textwidth}
        \centering
        \includegraphics[width=\textwidth]{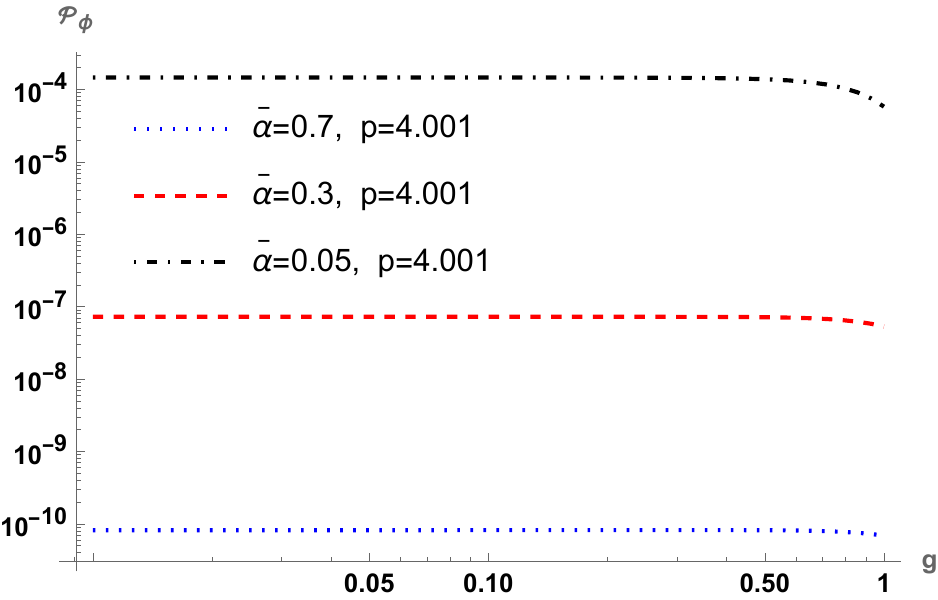}
    \end{subfigure}
    \caption{\small \it Variation of the spectral index and the power spectrum, respectively  \ref{spec}, \ref{P1}, with respect to the Yukawa coupling, \ref{wphi2}. The spectral index (power spectrum) increases (decreases) with respect to the Yukawa coupling.}
    \label{fig:both5}
\end{figure}

\noindent
As we have stated in the preceding section, if the inflation lasts for sufficiently long time, we must have $k \ll Ha$ in \ref{spectrumfull}, towards the end of inflation.  The leading contribution to the power spectrum of the two point correlation function for the scalar field ($f(\phi)=\phi$) is given by
\begin{equation}
    \mathcal{P}^{(2)}_{\phi}\simeq \frac{2 f_{1}^{2}}{\pi}\Gamma\left(2-\frac{2E_{1}}{H}\right)\sin\frac{\pi E_{1}}{H}\left(\frac{k_{\star}}{k_{\rm end}}\right)^{2E_{1}/H}
    \label{P1}
\end{equation}
 where $Ha=k_{\rm end}$   corresponds to the CMB modes at  late times after the exit off the horizon and $k_{\star}$ is called the pivot scale. We will evaluate $\mathcal{P}_{\phi}$ at  $k_{\star}=0.05\text{Mpc}^{-1}$, as suggested by the {\it Planck Data} (2018)~\cite{Planck:2018jri, Planck:2019kim}. We will also take $k_{\text{end}}\sim10^{23}\text{Mpc}^{-1}$~\cite{Ebadi:2023xhq}. As $k_{\star}/k_{\rm end} \ll 1$, the lowest spectral coefficient appearing in \ref{P1} is $f_1$, which  can be computed numerically from \ref{coefficient}, \ref{eq:schrodinger equation}.  The spectral index reads
\begin{equation}
n_s=1+\frac{2E_1}{H}
\label{spec}
\end{equation}
We have plotted the variation of the spectral index and the power spectrum \ref{P1}, in \ref{fig:both1},   \ref{fig:both4} and \ref{fig:both5} with respect to different parameters. In particular, \ref{fig:both5} shows that the spectral index (power spectrum) increases (decreases) with respect to the Yukawa coupling. Also in all the cases, we see a blue tilted power spectrum for the spectator scalar. We have also provided some generic numerical values of the spectral coefficient $f_1$ in \ref{f1T}.\\

\begin{table}[h]
    \centering
    \begin{tabular}{|c|c|c|c|}
        \hline
        $\bar{\alpha}$ & $p$ & $g$ & $|f_1|$ \\
        \hline
        0.01  & 4.001  & 0.01  &0.81822  \\
        \hline
        0.01  & 4.001  & 0.1  & 0.81735  \\
        \hline
        0.01  & 4.001  & 0.5  & 0.79129  \\
        \hline
        0.05  & 4.001  & 0.01  & 0.54723 \\
        \hline
        0.3  & 4.001  & 0.01  & 0.34969  \\
        \hline
        0.7  & 4.001  & 0.01  & 0.28295  \\
        \hline
        0.01  & 4.001  & 0.01  & 0.81822  \\
        \hline
        0.01  & 4.3  & 0.01  & 0.78318  \\
        \hline
        0.01  & 4.5  & 0.01  & 0.76324  \\
        \hline
    \end{tabular}
    \caption{\small \it Some generic numerical values for the spectral coefficient $f_1$, \ref{coefficient}.}
    \label{f1T}
\end{table}

\begin{figure}[hbt!]
    \centering
    \begin{subfigure}[b]{0.45\textwidth}
        \centering
        \includegraphics[width=\textwidth]{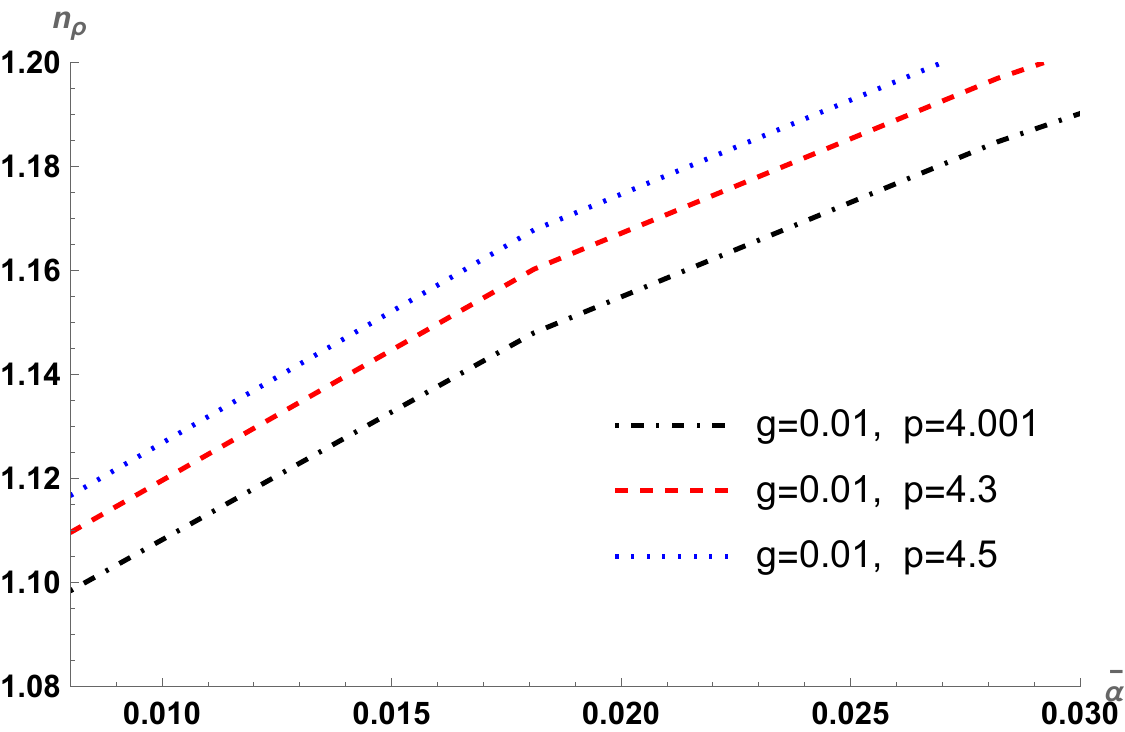}
    \end{subfigure}
    \hfill
    \begin{subfigure}[b]{0.45\textwidth}
        \centering
        \includegraphics[width=\textwidth]{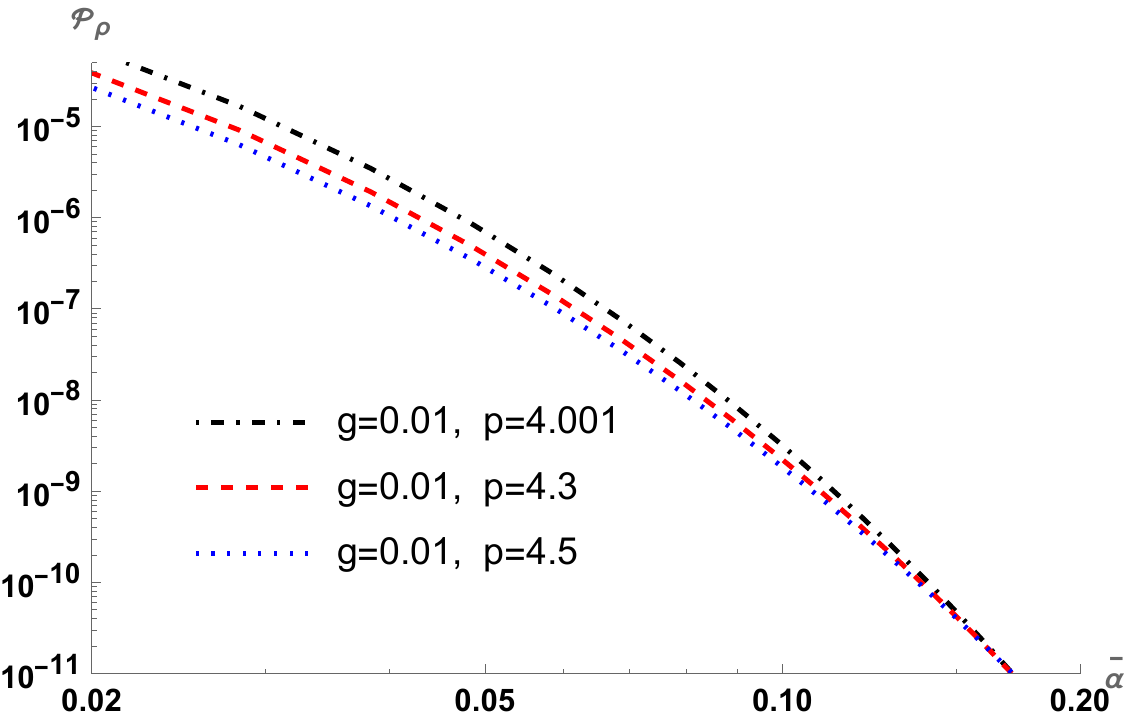}
    \end{subfigure}
    \caption{\small \it Variation of the spectral  index and the power spectrum   for the density contrast for the spectator scalar, \ref{spec2}, \ref{ps-dp},  with respect to the dimensionless parameter, $\bar{\alpha}$. The above figures also shows the variation with respect to the parameter $p$. }
    \label{fig:both9}
\end{figure}

\begin{figure}[hbt!]
    \centering
    \begin{subfigure}[b]{0.45\textwidth}
        \centering
        \includegraphics[width=\textwidth]{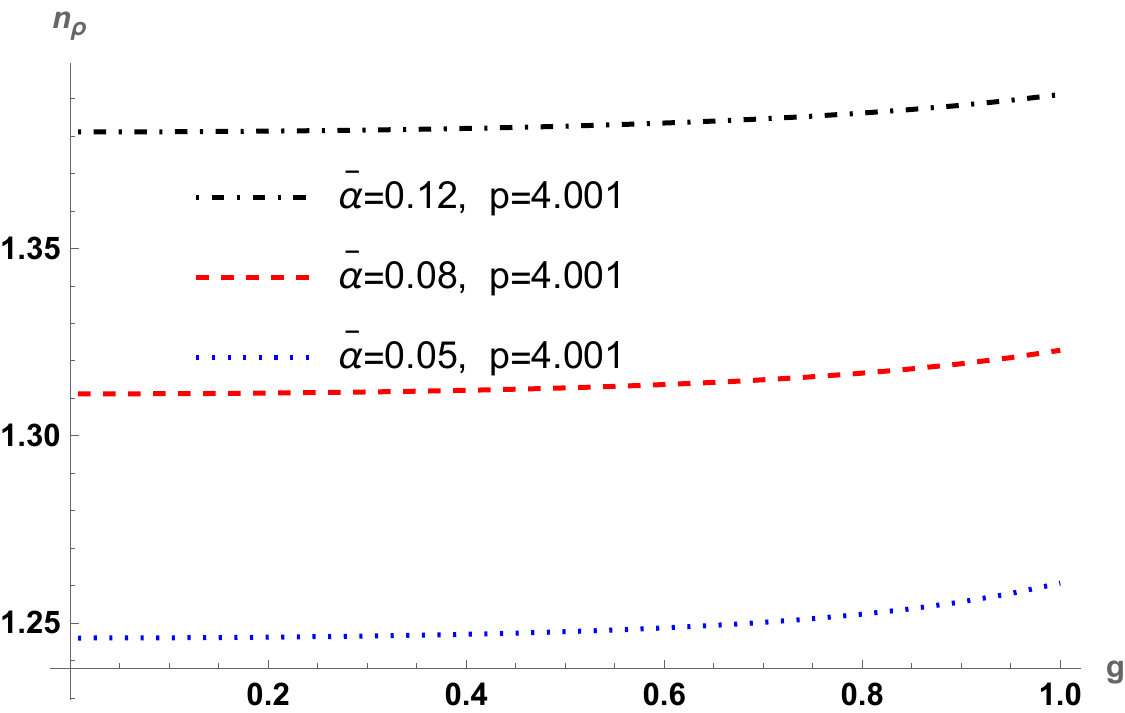}
    \end{subfigure}
    \hfill
    \begin{subfigure}[b]{0.45\textwidth}
        \centering
        \includegraphics[width=\textwidth]{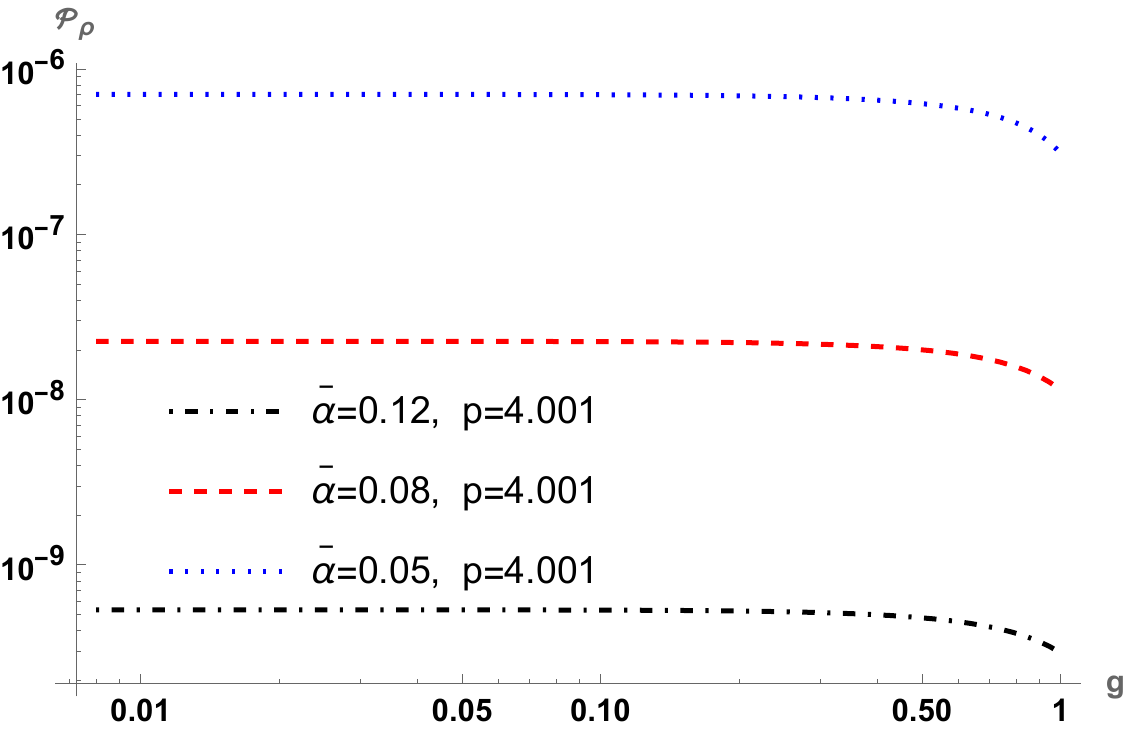}
    \end{subfigure}
    \caption{\small \it Variation of the spectral index and the power spectrum and  for the density contrast for the spectator scalar, \ref{spec2}, \ref{ps-dp},  with respect to the Yukawa coupling. }
    \label{fig:both10}
\end{figure}

\noindent
We now wish to compute the 
density contrast for the spectator scalar. Since we have assumed that the spectator is slowly moving, we take the energy density of the field to be   $\rho_{\phi}\simeq 8\pi GV_{\text{eff}}(\phi)$. One defines the variance as a measure of fluctuation of this density as 
\begin{eqnarray}
\label{eq:density-perturbation}
\delta_{\phi} = \frac{\delta \rho_{\phi}}{\langle\rho_{\phi}\rangle}&:=\frac{V_{\text{eff}}(\phi)-\langle V_{\text{eff}}(\phi)\rangle}{\langle V_{\text{eff}}(\phi)\rangle} 
\end{eqnarray}
 What should be the correlation between two $\delta_{\phi}$'s at equal time? 
Since  $V_{\text{eff}}(\phi)$ is even in $\phi$, the density contrast $\delta_{\phi}$ will also be even. Thus for the  leading nontrivial contribution, we must set $n=2$ in \ref{equ:timetospace}, to have the power spectrum 
\begin{equation}\label{ps-dp}
    \mathcal{P}^{(2)}_{\rho}\simeq\frac{2 f_{2}^{2}}{\pi}\Gamma\left(2-\frac{2E_{2}}{H}\right)\sin\left(\frac{\pi E_{2}}{H}\right)\left(\frac{k_{\star}}{k_{\rm end}}\right)^{2E_{2}/H}
\end{equation}
and the corresponding spectral index is given by
\begin{equation}
    n_{\rho}=1+\frac{2E_{2}}{H}
    \label{spec2}
\end{equation}
We have plotted the variations of \ref{ps-dp} and \ref{spec2} in \ref{fig:both9} and \ref{fig:both10}, after evaluating $f_2$ numerically. Once again, we see blue tilted spectra. We have also provided some numerical values for the spectral coefficient $f_2$ in \ref{f2T}.\\
\begin{table}[h]
    \centering
    \begin{tabular}{|c|c|c|c|}
        \hline
        $\bar{\alpha}$ & $p$ & $g$ & $|f_2|$ \\
        \hline
        0.001  & 4.001  & 0.001  &0.00140  \\
        \hline
        0.001  & 4.001  & 0.03  & 0.00150  \\
        \hline
        0.001  & 4.001  & 0.1  & 0.00240  \\
        \hline
        0.0001  & 4.001  & 0.01  & 0.00144  \\
        \hline
        0.001  & 4.001  & 0.01  & 0.00140  \\
        \hline
        0.02  & 4.001  & 0.01  & 0.00141  \\
        \hline
        0.001  & 4.001  & 0.01  & 0.00140 \\
        \hline
        0.001  & 4.3  & 0.01  & 0.00129  \\
        \hline
        0.001  & 4.5  & 0.01  & 0.00219  \\
        \hline
    \end{tabular}
    \caption{\small \it Some generic numerical values  for the spectral coefficient $f_2$, \ref{coefficient}.}
    \label{f2T}
\end{table}
%
\section{Three point correlation function}\label{S5}

We next wish to compute the three-point correlation function for the density fluctuation discussed above. The large scale structure of the universe, like that of the  clusters, superclusters, and galaxies are the result of the inhomogeneities and anisotropies of the primordial fluctuations that got stretched to super-Hubble scales at sufficiently late times and then re-entered after the end of  inflation. To capture these anisotropies in the spectrum, one needs to compute the bispectrum or the three point correlation function,  and then evaluate the measure of the non-linearity parameter $f_{\rm NL}$, which tells the amount of deviation from the usual Gaussian nature of the fluctuations. We shall explicitly evaluate the correlation function and associated non-Gaussianity for the density fluctuations, $\delta_{\phi}$, discussed in the preceding section.

We shall sketch the derivation of the three point function below, following e.g.~\cite{Motohashi:2012bb, Antoniadis:2011ib,Kehagias:2012pd} (also references therein).  From the isometries of the de Sitter space, the general expression for the three-point correlator is written as (see~\ref{appendix:a} for a brief sketch of the derivation),
\begin{eqnarray}
 &&   \xi_{f}^{(3)}(\vec{x}_{1},\vec{x}_{2},\vec{x}_3;t)= \langle f(\phi(\vec{x}_1,t))f(\phi((\vec{x}_2,t))f(\phi(\vec{x}_3,t))\rangle \nonumber\\ &&
=\mathcal{N}^{2}\sum_{a,b,c}f_{a}f_{b}f_{c}B_{abc}(HR_{12})^{-\frac{E_{a}+E_{b}-E_{c}}{H}}(HR_{23})^{-\frac{E_{b}+E_{c}-E_{a}}{H}}(HR_{31})^{-\frac{E_{c}+E_{a}-E_{b}}{H}} 
    \label{3p1}
     \end{eqnarray}
%
%
%
%
where $R_{ij}=a(t)|\vec{x}_i-\vec{x}_j|$. The coefficients $\mathcal{N}$ and $B_{abc}$ are given by
\begin{eqnarray}
\mathcal{N}^{-2}=\int d\phi \ e^{-2\nu(\phi)}, {\hspace{0.5cm}}
B_{abc}=\int d\phi \ e^{\nu(\phi)}\psi_a(\phi)\psi_b(\phi)\psi_{c}(\phi),
\label{norm}
\end{eqnarray}
and $\nu(\phi)$ and the spectral coefficients $f_{n}$ are respectively given by  \ref{effective potential}. Denoting now $w_{i}=(E_{a}+E_{b}-E_{c})/H$, $w_{j}=(E_{b}+E_{c}-E_{a})/H$ and $w_{k}=(E_{c}+E_{a}-E_{b})/H$, we have the three point correlation function in momentum space (e.g.~\cite{Antoniadis:2011ib,Kehagias:2012pd}), 
\begin{equation}
    \xi^{(3)}_{f}(\vec{k}_1,\vec{k}_2,\vec{k}_3;a)=\sum_{a,b,c}C_{abc}\int d^{3}\vec{x}_1 d^{3}\vec{x}_2 d^{3}\vec{x}_3 \frac{e^{i(\vec{k}_1\cdot\vec{x}_1+\vec{k}_2\cdot\vec{x}_2+\vec{k}_3\cdot\vec{x}_3)}}{(aH)^{w_{i}+w_{j}+w_{k}}} \frac{1}{|\vec{x}_1-\vec{x}_2|^{w_i} |\vec{x}_2-\vec{x}_3|^{w_j} |\vec{x}_3-\vec{x}_1|^{w_k}} 
    \label{3p3}
\end{equation}
where we have abbreviated
\begin{equation}
    C_{abc}=\mathcal{N}^{2} B_{abc}f_{a}f_{b}f_{c} \qquad ({\rm no~sum})
    \label{abc}
\end{equation}
Writing now, $\xi_f^{(3)}(\vec{k}_1,\vec{k}_2,\vec{k}_3;a)=(2\pi)^3\delta^{(3)}(\vec{k}_1+\vec{k}_2+\vec{k}_3)B_{S}(\vec{k}_{1},\vec{k}_{2}, \vec{k}_{3};a)$,  one derives from the above (\cite{Motohashi:2012bb, Antoniadis:2011ib,Kehagias:2012pd}, sketched in \ref{appendix:b})
\begin{equation}\label{bi-s3}
\begin{split}
B_{S}(\vec{k}_1,\vec{k}_2,\vec{k}_3;a)&=\sum_{a,b,c}\frac{C_{abc}}{(aH)^{w_{i}+w_{j}+w_{k}}}2^{7-w_{i}+w_{j}+w_{k}}\pi^{5/2}\frac{\Gamma(3-\frac{w_{i}+w_{j}+w_{k}}{2})\Gamma(\frac{3-w_k}{2})}{\Gamma(\frac{w_i}{2})\Gamma(\frac{w_j}{2})}\\
&\times k_{1}^{w_{i}+w_{j}+w_{k}-6}\int_{0}^{1}ds\frac{(1-s)^{\frac{w_j}{2}-1}s^{\frac{w_i}{2}-1}}{{\left[(1-s)X+sY\right]}^{\frac{3-{w_k}}{2}}} {}_2F_{1}\left( \frac{w_{i}+w_{j}+w_{k}}{2}-\frac{3}{2},\frac{3-w_{k}}{2},\frac{3}{2},{Z}\right)
\end{split}
\end{equation}
where $X$, $Y$ and $Z$ are given by \ref{XYZ}.
 We now wish to evaluate the above for $w_{i}=w_{j}=w_{k}=w~({\rm say})$, known as the scaling parameter, and in the squeezed limit.  In this limit,
one studies  the correlation between one super-Hubble with  two sub-horizon modes, for which  we have $k_1 \ll k_2\sim k_3$, and  $X\sim Y\ge1$.  This will simplify \ref{bi-s3} greatly.
%

\subsection{Numerical analyses and  the squeezed limit and non-Gaussianity}\label{BSS}

\noindent
Defining  $z_1=k_1/k_3$ and $z_2=k_2/k_3$, we now introduce the {\it shape function}, $S(z_1,z_2,w)$, in terms of which and the scaling parameter the simplified  bispectrum reads
\begin{equation}\label{sqeezed-bs5}
    B_{S}^{\rm sq}(\vec{k}_1,\vec{k}_2,\vec{k}_3)=\left(\frac{k_{3}}{aH}\right)^{3w}\frac{1}{(k_1k_2k_3)^2}S(z_1,z_2,w)
\end{equation}
where the shape function explicitly reads
\begin{equation}
    S(z_1,z_2,w)=\sum_{a,b,c=1}^3\gamma_{\rm sq}\left[C_{abc}z_{1}^{2w-1}z_2^{w-1}+C_{bca}z_1^{{2}}z_2^{{2w-1}}+C_{cba}z_{1}^{w-1}z_2^{2}\right]
\end{equation}
where we have abbreviated 
\begin{equation}
    \gamma_{\rm sq}=(2\pi)^3 \frac{\Gamma(\frac{3}{2}-\frac{w}{2})\Gamma(\frac{3}{2}- 2w)}{2^{3(w-1)}\Gamma(\frac{w}{2})\Gamma(w)}.
    \end{equation}
For $f(\phi)$ to be the density fluctuation $\delta_{\phi}$, \ref{eq:density-perturbation}, we are interested in, we see that for the coefficient $B_{abc}$, the leading non-trivial contribution comes from $B_{222}$, while the spectral coefficient $f_2$ can be read off from \ref{coefficient}. Also, the relevant scaling parameter in this case reads, $w=E_2/H$. Putting things together, the leading expression for the shape function for the bispectrum for the density perturbations is given by
\begin{equation}\label{eq:shapefunction}
S(z_1,z_2,w)\simeq \mathcal{N}^{2}B_{222}(f_{2}^{2})^{3}\gamma_{\rm sq}\left[z_{1}^{2w-1}z_2^{w-1}+z_1^{{2}}z_2^{{2w-1}}+z_{1}^{w-1}z_2^{2}\right]
\end{equation}
which we wish to evaluate numerically.

Note that by the virtue of the  spatial translational invariance of the de Sitter background, the momentum space  \ref{eq:3pt-bs} implies that $\vec{k}_1+\vec{k}_2+\vec{k}_3=0$. Clearly, this constraint leads to the possibility of a large number of shape functions~\cite{Cabass:2022avo, Komatsu:2001rj, yadav2010primordial}. For example,  the {\it equilateral} shape corresponds to, $k_1\approx k_2\approx k_3$,  e.g.~\cite{Gwyn:2012pb}. The {\it folded} shape function for which $k_1\approx 2k_2\approx 2k_3$, might be particularly interesting in the case when modes do not correspond to the standard Bunch-Davies vacuum, e.g.~\cite{Meerburg:2009ys}. 
In particular as we have stated earlier, in the squeezed limit we are interested in, we have $z_1=k_1/k_3 \to 0$ and $z_2=k_2/k_3 \to 1$ in \ref{eq:shapefunction}, corresponding  to one super-Hubble and two sub-Hubble modes.    This plays a crucial role in understanding the nature of the primordial non-Gaussianities we wish to eventually compute.

\begin{figure}[htbp]
    \centering
    \begin{subfigure}[b]{0.3\textwidth}
        \includegraphics[width=\textwidth]{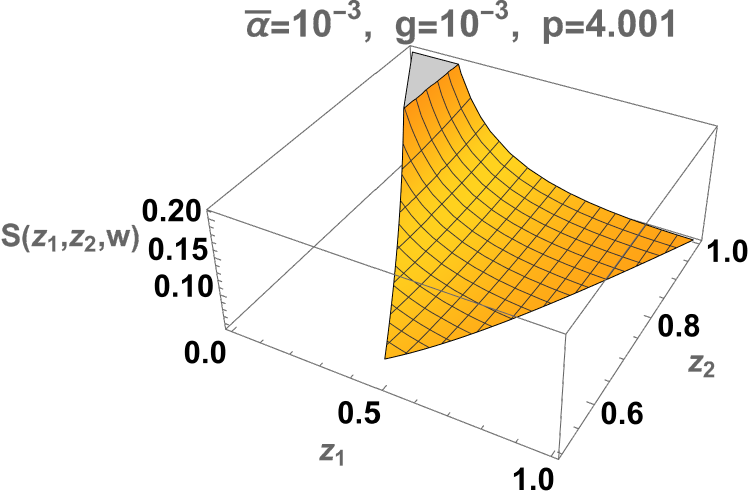}
        \caption{}
        \label{}
    \end{subfigure}
    \hfill
    \begin{subfigure}[b]{0.3\textwidth}
        \includegraphics[width=\textwidth]{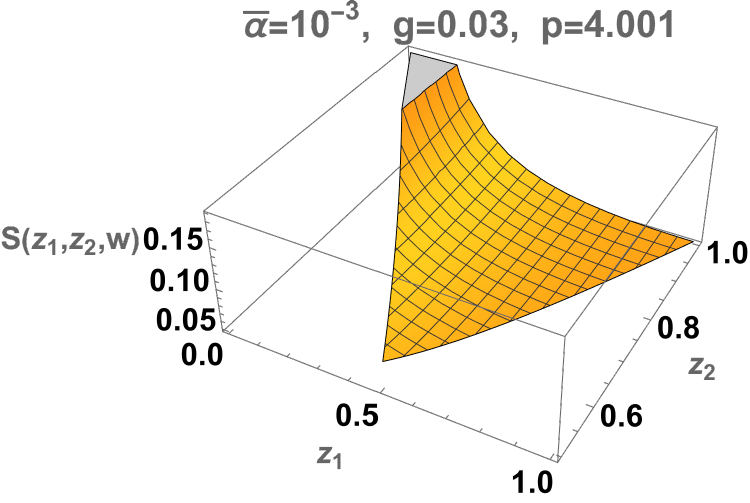}
        \caption{}
        \label{}
    \end{subfigure}
    \hfill
    \begin{subfigure}[b]{0.3\textwidth}
        \includegraphics[width=\textwidth]{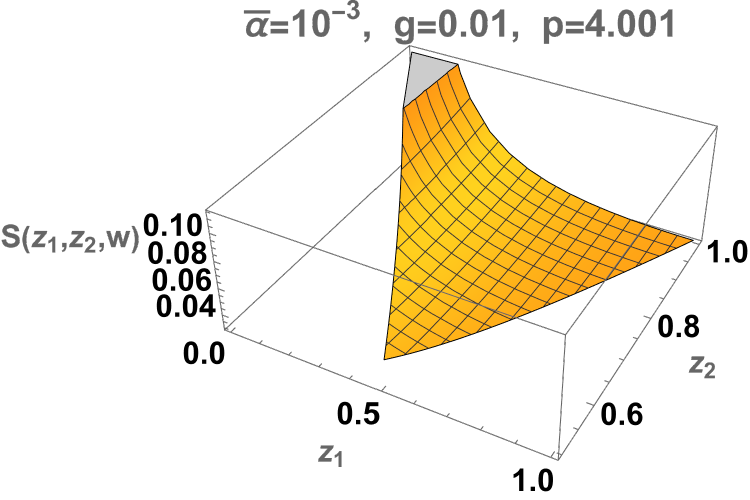}
        \caption{}
        \label{}
    \end{subfigure}
    \caption{\small \it The bispectrum shape function $S(z_1, z_2, w)$,  \ref{eq:shapefunction}, for different values of $g$. The parameters $\bar{\alpha}$ and $p$ are held  fixed. The shape function is plotted as a function of $z_1 = k_1 / k_3$ and $z_2 = k_2 / k_3$. All plots are normalised in the equilateral limit ($z_1 = z_2$), and the region $1 - z_2 \leq z_1 \leq z_2$ is set to zero. The shape function peaks at the squeezing limit, $z_1\to 0$ and $z_2\to 1$. See main text for discussion. }
    \label{fig:plot8}
\end{figure}

\begin{figure}[htbp]
    \centering
    \begin{subfigure}[b]{0.3\textwidth}
        \includegraphics[width=\textwidth]{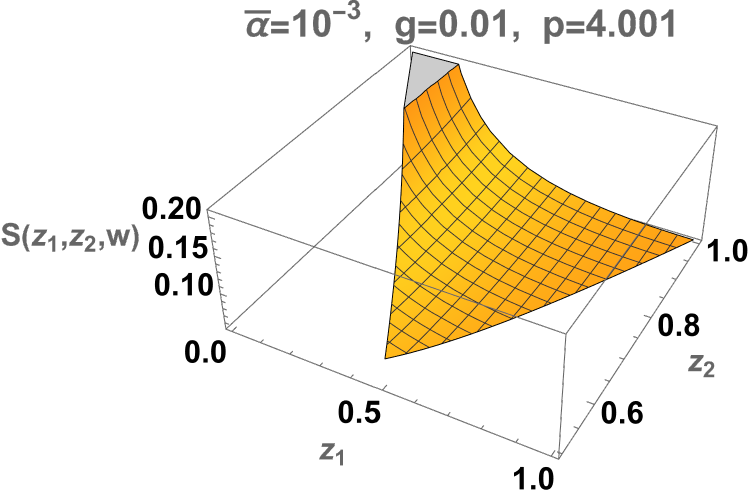}
        \caption{}
        \label{}
    \end{subfigure}
    \hfill
    \begin{subfigure}[b]{0.3\textwidth}
        \includegraphics[width=\textwidth]{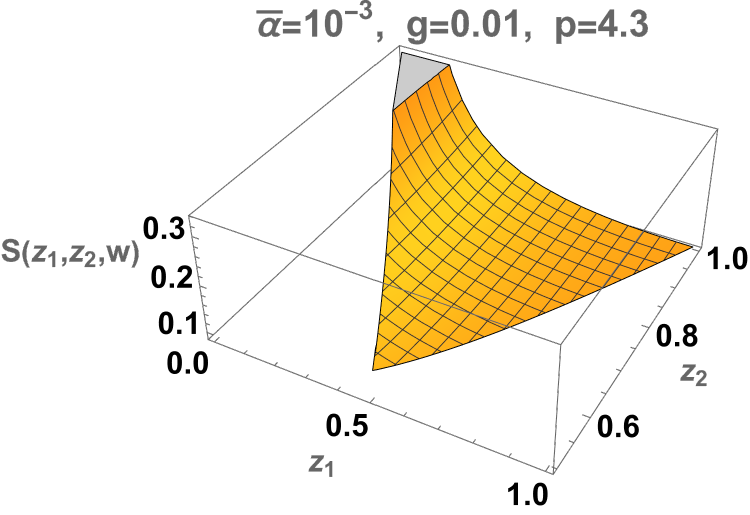}
        \caption{}
        \label{}
    \end{subfigure}
    \hfill
    \begin{subfigure}[b]{0.3\textwidth}
        \includegraphics[width=\textwidth]{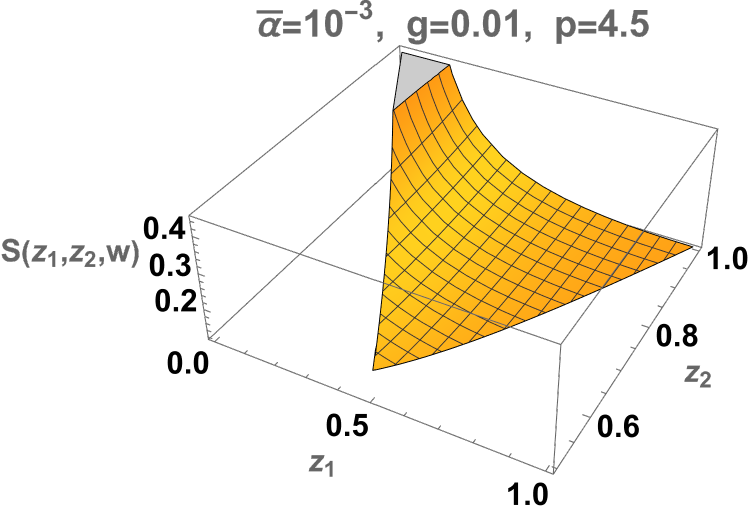}
        \caption{}
        \label{}
    \end{subfigure}
  \caption{\small \it The bispectrum shape function $S(z_1, z_2, w)$,  \ref{eq:shapefunction}, for different values of the effective potential parameter $p$. The parameters $\bar{\alpha}$ and $g$ are held fixed. The shape function is plotted as a function of $z_1 = k_1 / k_3$ and $z_2 = k_2 / k_3$. All plots are normalised in the equilateral limit ($z_1 = z_2$), and the region $1 - z_2 \le z_1 \le z_2$ is set to zero. See main text for discussion.}
\label{fig:plot9}
\end{figure}

\begin{figure}[htbp]
    \centering
    \begin{subfigure}[b]{0.3\textwidth}
        \includegraphics[width=\textwidth]{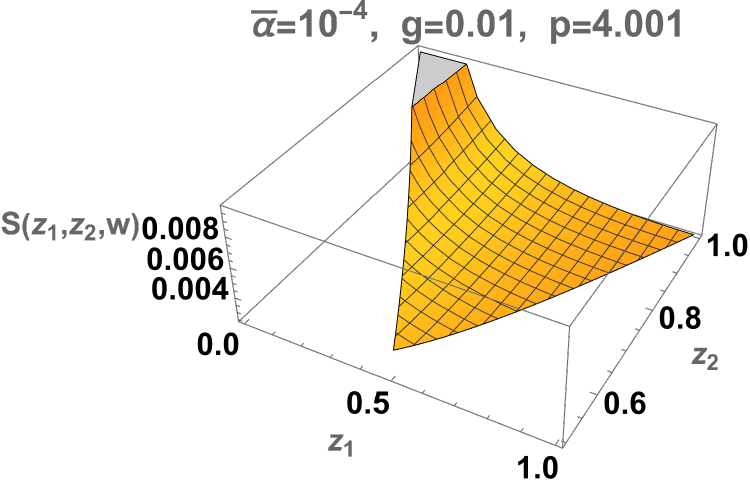}
        \caption{}
        \label{}
    \end{subfigure}
    \hfill
    \begin{subfigure}[b]{0.3\textwidth}
        \includegraphics[width=\textwidth]{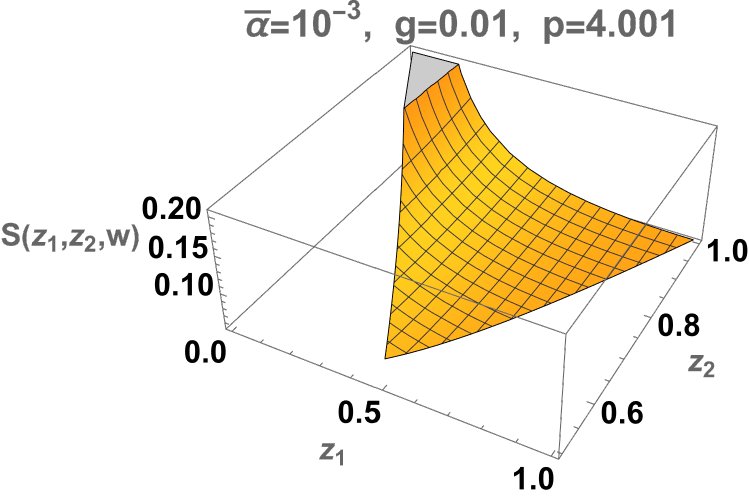}
        \caption{}
        \label{}
    \end{subfigure}
    \hfill
    \begin{subfigure}[b]{0.3\textwidth}
        \includegraphics[width=\textwidth]{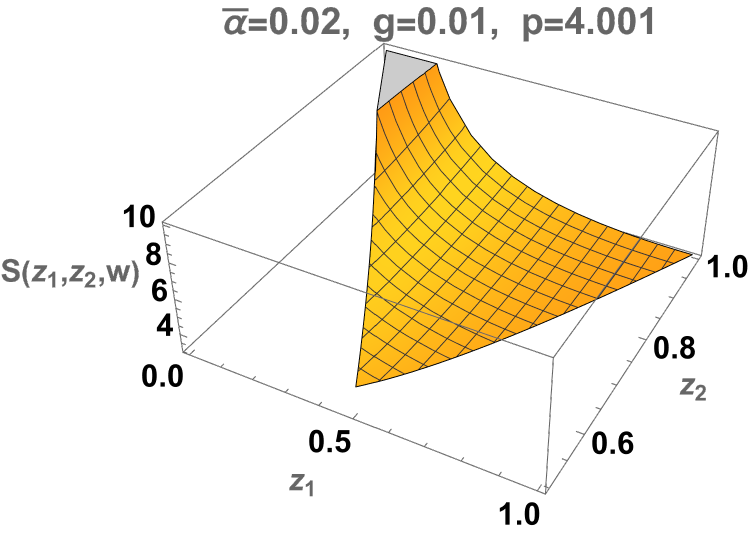}
        \caption{}
        \label{}
    \end{subfigure}
     \caption{\small \it The bispectrum shape function $S(z_1, z_2, w)$,  \ref{eq:shapefunction}, for different values of the effective potential parameter $\bar{\alpha}$, while $g$ and $p$ are held.  All plots are normalised in the equilateral limit ($z_1 = z_2$), and the region $1 - z_2 \le z_1 \le z_2$ is set to zero. See main text for discussion.}
    \label{fig:plot10}
\end{figure}

 In \ref{fig:plot8}, we have plotted the shape function $S(z_1,z_2,w)$, \ref{eq:shapefunction}, with respect to $z_1$ and $z_2$ with different values  of the Yukawa coupling, $g$, while the parameters $\bar{\alpha}$ and $p$ of the effective potential are held fixed. We note that the shape function is peaked in the squeezed limit, $k_1\rightarrow0$ and $k_2\approx k_3$. We also note  that the shape function for lower $g$ values is more highly peaked compared to its larger values. Next in \ref{fig:plot9}, we have plotted the shape function for which $\bar{\alpha}$ and $g$ values are held fixed, and varied with respect to $z_1$, $z_2$, and the parameter $p$. We see that the shape function is highly peaked for higher $p$ values, opposed to that of the Yukawa coupling. Likewise, \ref{fig:plot10} shows that the shape function is more highly peaked for larger $\bar{\alpha}$ values, similar to that of $p$ but contrary to that of $g$.  Thus it seems fair enough to conclude that the Yukawa coupling makes the shape function broader in the squeezed limit.        
\bigskip

\noindent
The  three point correlation function for the density fluctuation discussed above allows us to investigate the density perturbations which deviates from the usual Gaussian spectrum. At the level of the scalar field, the local non-Gaussianity   parameter  $f_{\rm NL}^{\rm loc}$, is defined as~\cite{Komatsu:2001rj},
\begin{equation}\label{eq:ng-1}
    \phi(x)=\phi_G(x)+\frac{3}{5}f_{\rm NL}^{\rm loc}(\phi_G^2(x)-\langle\phi_G(x)\rangle^2)
\end{equation}
where $\phi_G(x)$ is a Gaussian random field, with $\langle\phi_G(x)\rangle=0$. Now the bispectrum, \ref{sqeezed-bs5}, can also be written as~\cite{Boubekeur:2005fj}, 
\begin{equation}
    B_{S}^{\rm sq}(\vec{k}_1,\vec{k}_2,\vec{k}_3)=\frac{6 f_{\rm NL}^{\rm loc}}{5}\left[P^{(2)}_{\rho}(\vec{k}_1)P^{(2)}_{\rho}(\vec{k}_2)+P^{(2)}_{\rho}(\vec{k}_2)P^{(2)}_{\rho}(\vec{k}_3)+P^{(2)}_{\rho}(\vec{k}_3)P^{(2)}_{\rho}(\vec{k}_1)\right]
    \label{BS}
\end{equation}
where $P^{(2)}_{\rho}(\vec{k})$ is related to the two-point correlation function corresponding to the density perturbation ${\cal P}^{(2)}_{\rho}(\vec{k})$ as, $P^{(2)}_{\rho}(\vec{k})={\cal P}^{(2)}_{\rho}(\vec{k})/k^3$,  \ref{ps-dp}. 

\begin{figure}[htbp]
    \centering
    \begin{subfigure}[b]{0.3\textwidth}
        \includegraphics[width=\textwidth]{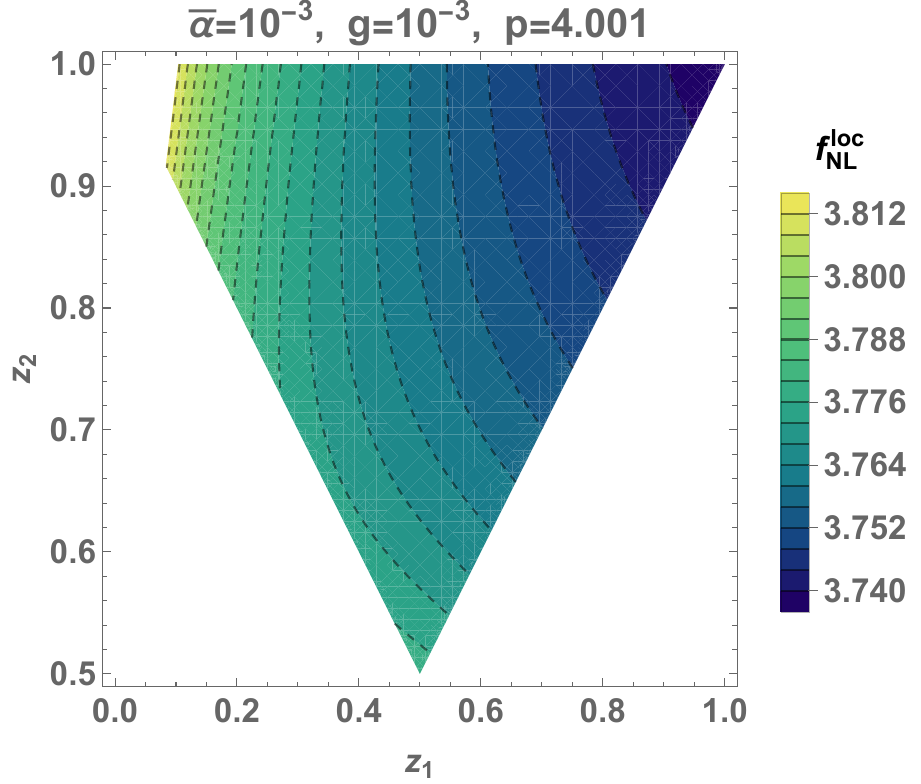}
        \caption{}
        \label{}
    \end{subfigure}
    \hfill
    \begin{subfigure}[b]{0.3\textwidth}
        \includegraphics[width=\textwidth]{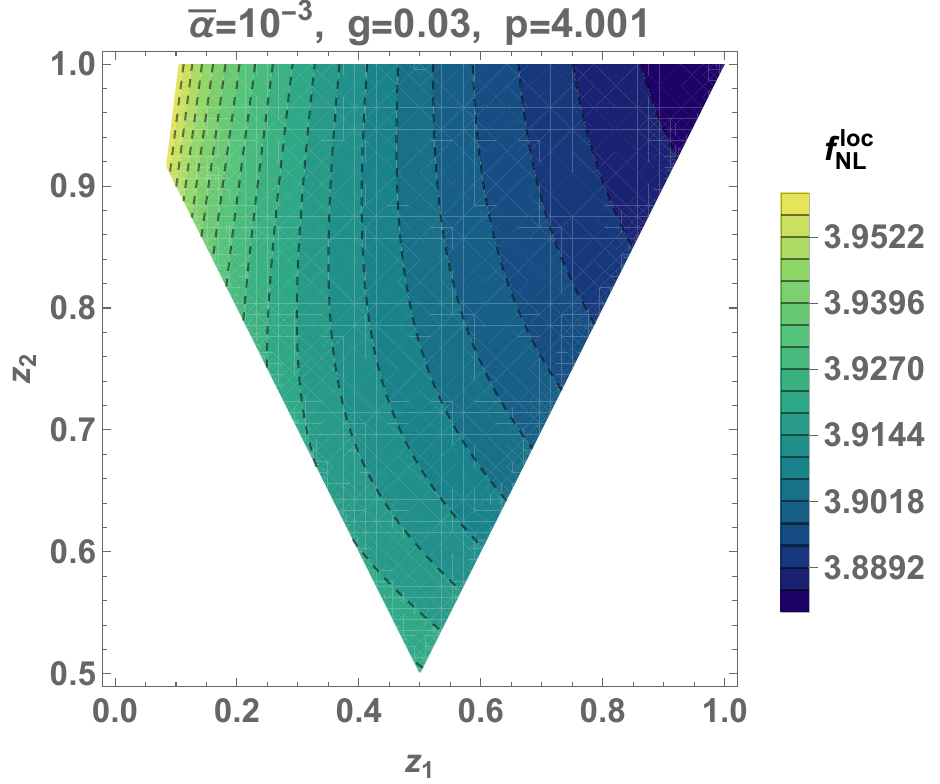}
        \caption{}
        \label{}
    \end{subfigure}
    \hfill
    \begin{subfigure}[b]{0.3\textwidth}
        \includegraphics[width=\textwidth]{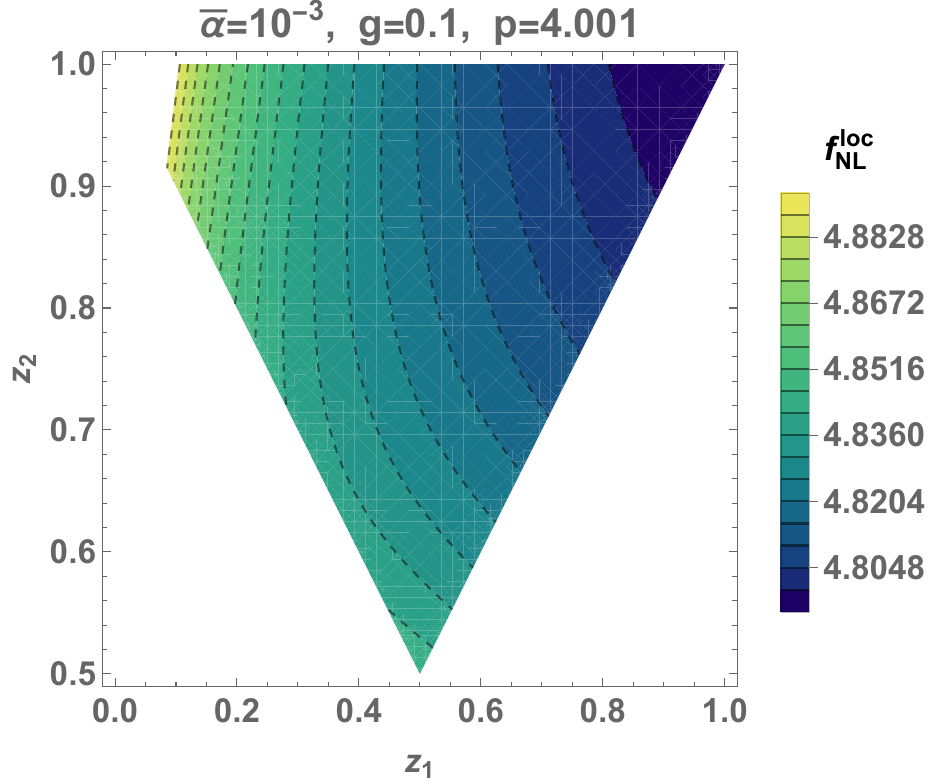}
        \caption{}
        \label{fig:plot3}
    \end{subfigure}
     \caption{\small \it Contour plots showing the variation of 
 the non-Gaussianity parameter $f_{\rm NL}^{\rm loc}$,\ref{eq:fnl-local}, with respect to $z_1=k_1/k_3$ and $z_2=k_2/k_3$. The parameters $\bar{\alpha}$ and $p$ are held fixed in all cases.  All plots are normalised in the equilateral limit ($z_1 = z_2$), and the region $1 - z_2 \le z_1 \le z_2$ is set to zero. These plots show increase of the non-Gaussianity in the squeezed limit with increasing Yukawa coupling. See main text for discussion.}
    \label{fig:plot11}
\end{figure}

\begin{figure}[htbp]
    \centering
    \begin{subfigure}[b]{0.3\textwidth}
        \includegraphics[width=\textwidth]{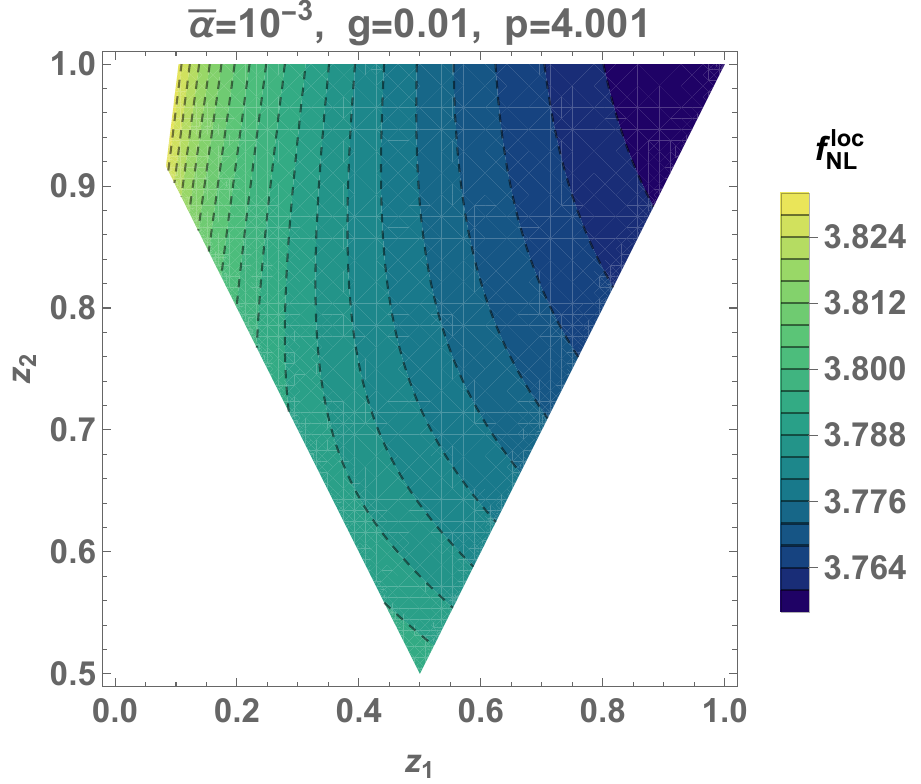}
        \caption{}
        \label{}
    \end{subfigure}
    \hfill
    \begin{subfigure}[b]{0.3\textwidth}
        \includegraphics[width=\textwidth]{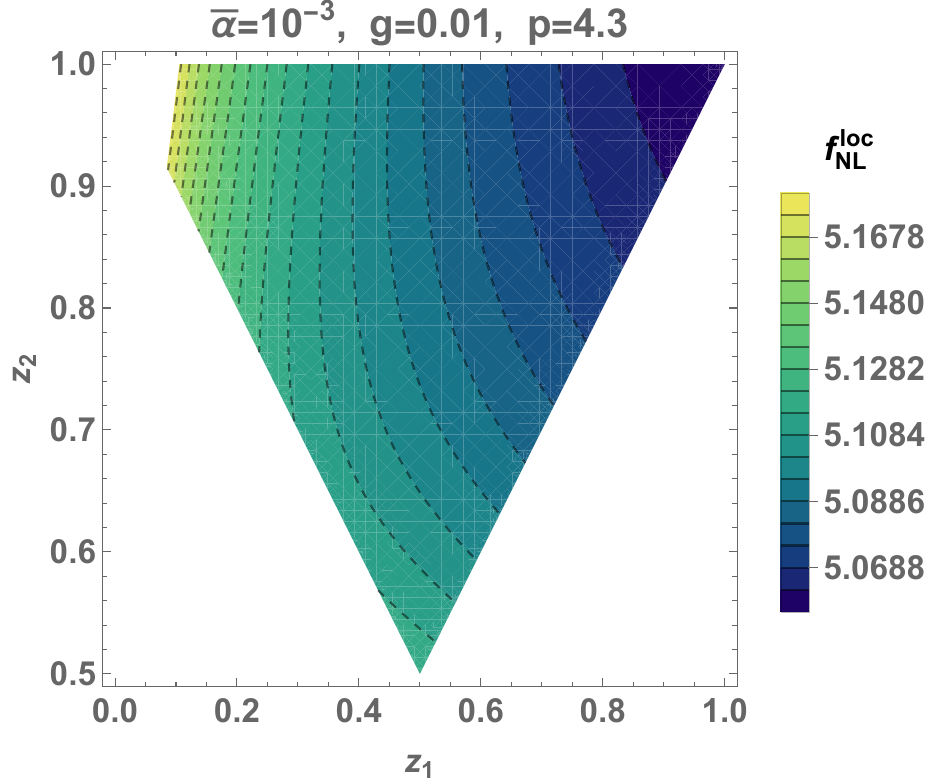}
        \caption{}
        \label{}
    \end{subfigure}
    \hfill
    \begin{subfigure}[b]{0.3\textwidth}
        \includegraphics[width=\textwidth]{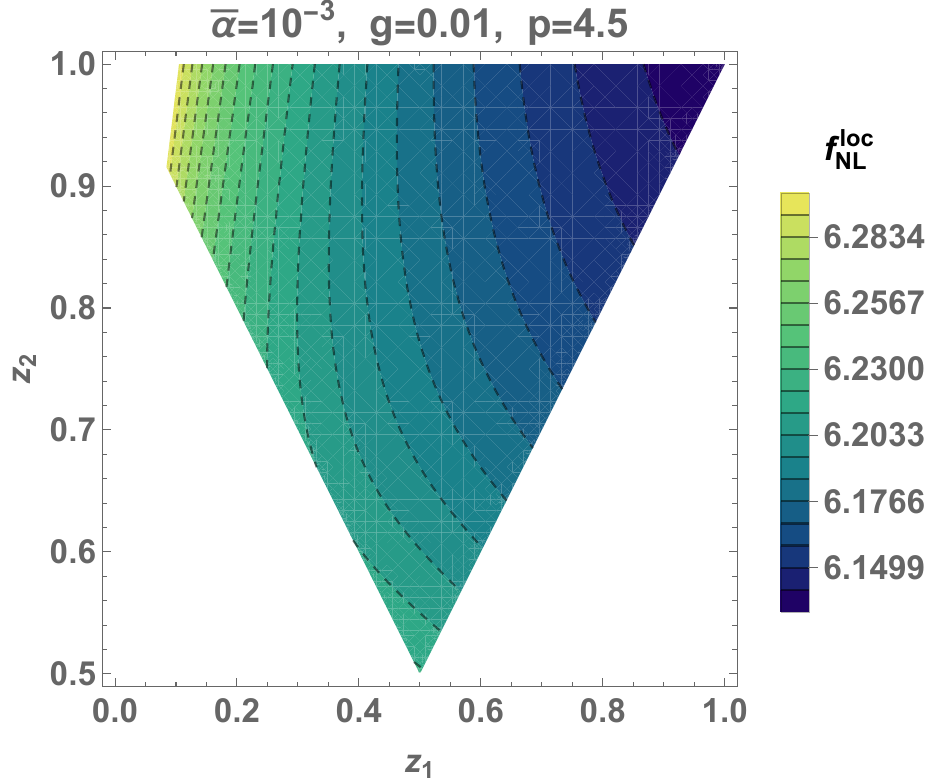}
        \caption{pp2}
        \label{}
    \end{subfigure}
    \caption{\small \it Contour plots showing the variation of 
$f_{\rm NL}^{\rm loc}$, \ref{eq:fnl-local}, with respect to $z_1=k_1/k_3$ and $z_2=k_2/k_3$ for different values of the parameter $p$. The parameters $\bar{\alpha}$ and $g$ are held fixed in all cases.  All plots are normalised in the equilateral limit ($z_1 = z_2$), and the region $1 - z_2 \le z_1 \le z_2$ is set to zero.}
    \label{fig:plot12}
\end{figure}

\begin{figure}[htbp]
    \centering
    \begin{subfigure}[b]{0.3\textwidth}
        \includegraphics[width=\textwidth]{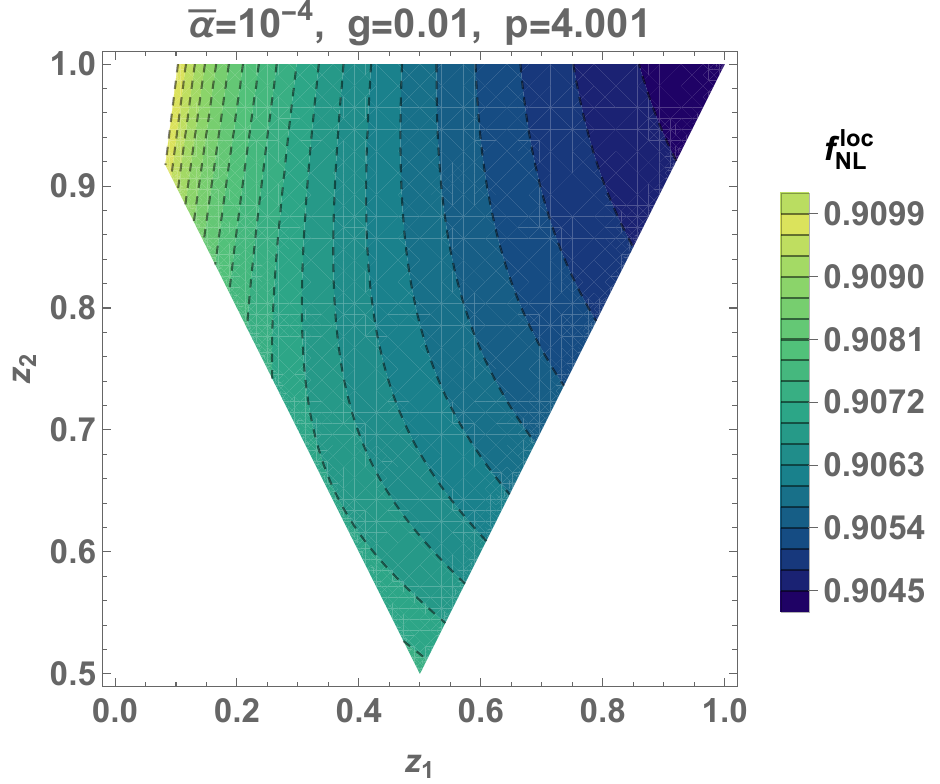}
        \caption{}
        \label{}
    \end{subfigure}
    \hfill
    \begin{subfigure}[b]{0.3\textwidth}
        \includegraphics[width=\textwidth]{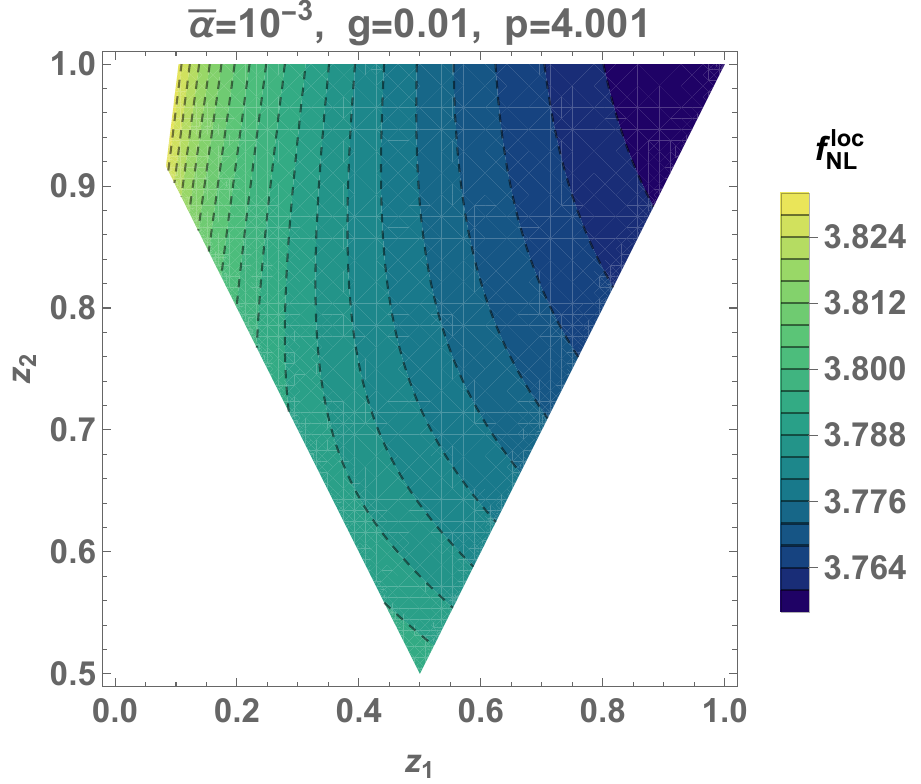}
        \caption{}
        \label{}
    \end{subfigure}
    \hfill
    \begin{subfigure}[b]{0.3\textwidth}
        \includegraphics[width=\textwidth]{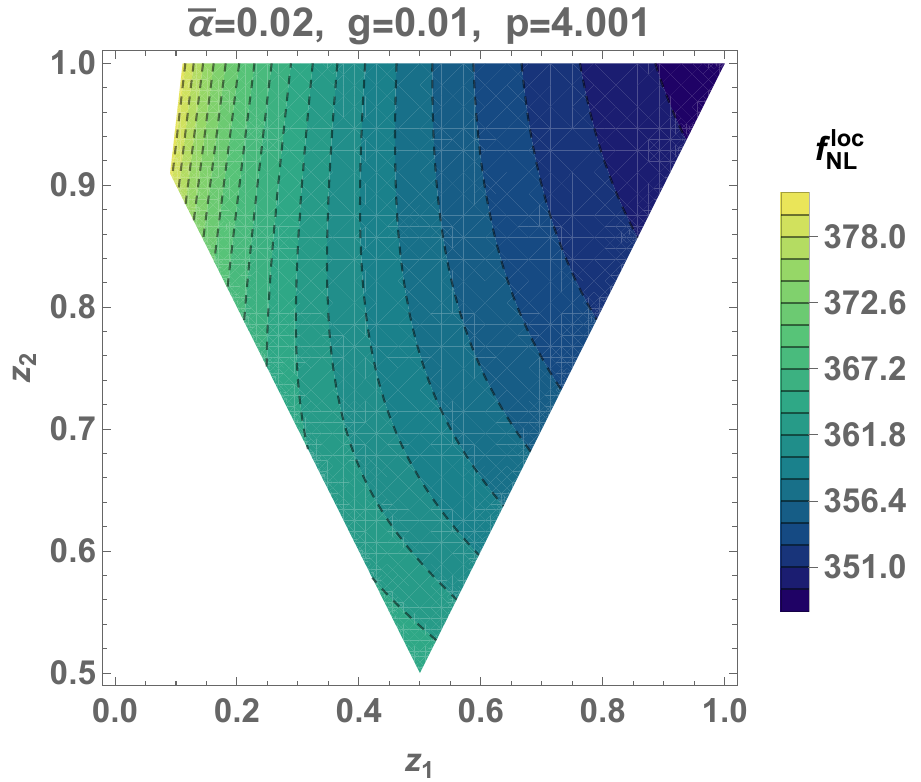}
        \caption{}
        \label{}
    \end{subfigure}
     \caption{\small \it Contour plots showing the variation of 
$f_{\rm NL}^{\rm loc}$, \ref{eq:fnl-local}, with respect to $z_1=k_1/k_3$ and $z_2=k_2/k_3$ for different values of $\bar{\alpha}$. The parameters $g$ and $p$ are held fixed.  All plots are normalised in the equilateral limit ($z_1 = z_2$), and the region $1 - z_2 \le z_1 \le z_2$ is set to zero.}
    \label{fig:plot13}
\end{figure}

We now Combine \ref{sqeezed-bs5}, \ref{BS} and \ref{ps-dp}, to arrive after a little algebra at an expression for the non-Gaussianity parameter suitable for our present purpose
\begin{equation}\label{eq:fnl-local}
    f_{\rm NL}^{\rm loc}=\frac{5}{6}\left(\frac{k_3}{aH}\right)^{-w} \frac{S(z_1,z_2,w)}{\Delta_{\phi}^{2}\left[z_{1}^{2w-1}z_2^{2w-1}+z_1^{{2}}z_2^{{2w-1}}+z_{1}^{2w-1}z_2^{2}\right]}
\end{equation}
where
\begin{equation}
    \Delta_{\phi}=\frac{2 f_{2}^{2}}{\pi}\Gamma\left(2-\frac{2E_{2}}{H}\right)\sin\left(\frac{\pi E_{2}}{H}\right)
\end{equation}
We shall evaluate $f_{\rm NL}^{\rm loc}$ at the pivot scale $k_3=k_{\star}=0.05\text{Mpc}^{-1}$, and at  $k_{\text{end}}=a_{\text{end}}H_{\text{end}}\sim10^{23}\text{Mpc}^{-1}$, as earlier. In \ref{fig:plot11}, we have plotted $f_{\rm NL}^{\rm loc}$ with respect to $z_1$ and $z_2$ and the Yukawa coupling $g$, while  $\bar{\alpha}$ and $p$ are held fixed. We may note here constraint from the  Planck Data (2018)~\cite{Planck:2019kim}
\begin{equation}\label{eq:fnlconstraint}
    -5.9\le f_{\rm NL}^{\rm loc}\le 4.1 \hspace{1cm} (\text{at} \hspace{0.2cm} 68\%),
\end{equation}
If we apply this constraint  for our spectator field as well, 
 \ref{fig:plot11} shows that in the squeezed limit, $f_{\rm NL}^{\rm loc}$ goes out of the upper bound for higher values of the Yukawa coupling, $g$. Likewise, \ref{fig:plot12} and \ref{fig:plot13} show the variation of the non-Gaussianity parameter with respect to the other parameters of the effective potential. We conclude that all of them increase the non-Gaussianity with their increasing values.
%

\section{Conclusion}\label{S6}
In this work we have investigated the effect of the Yukawa coupling on the two and three point correlation functions for a stochastic spectator scalar field. We have taken the one loop effective action for the fermion derived earlier in~\cite{Miao:2006pn}. We have chosen the scalar field's potential to be generic quintessence-like~\ref{scpot}, so that the total effective potential~\ref{eq:eff-pot}, is bounded from below and admits late time equilibrium state for the spectator. Using this total effective potential, we have numerically investigated the correlation functions using the stochastic spectral expansion method~\cite{Starobinsky:1994bd,  Markkanen:2019kpv, Markkanen:2020bfc}, respectively in \ref{S4} and \ref{S5}. For the two point correlation function, we have seen that the Yukawa coupling $g$ pushes the power spectrum and the spectral index towards blue side. This is qualitatively similar to the case of quartic self interaction reported in~\cite{Markkanen:2019kpv}. For the three point correlation, we have shown that with increasing $g$-value, the peak of the shape function gets flattened in the squeezed limit. The local non-Gaussianity parameter also increases with increasing $g$. The results do not depend upon the sign of $g$.

Note that the fermionic part of the effective action we have taken is {\it exact} and not a local, ultraviolet one,  for it uses the exact massless fermion propagator in de Sitter. Such UV limit corresponds to the high field limit ($\phi/H \gg 1$) in~\ref{eq:eff-pot}~\cite{Miao:2006pn}. Inclusion of a fermion rest mass would  correspond to the change : $g \phi \to m+g\phi$ in the effective action. This will make the effective potential asymmetric in $\phi$, unlike the present massless case.  However, it is easy to note from  \ref{wphi2}  that even after  this change for a massive fermion, there will remain a reflection symmetry of the effective potential around $\phi_0=- m/g$, although as a whole it is asymmetric. While we may expect qualitative new effects stemming from the fermion mass,  it seems reasonable to anticipate that for a very light  fermion ($m/H\ll 1$),  the chief features of the results found here would remain by and large intact.

 Finally, what will happen if we take the scalar to be an inflaton instead of a spectator? In this case we need to work in  the quasi-de Sitter background along with various slow roll parameters.  In particular for a massive fermion with mass proportional to the Hubble rate, an exact propagator  exists if the principal slow roll parameter is constant~\cite{Koksma:2009tc}. This propagator can be used to derive a modified  effective action.   A second improvement will be to study the correlations generated by the curvature   perturbation by the inflaton.  There are standard ways to compute such correlations, such as the  stochastic $\delta N$ formalism~\cite{Abolhasani:2011yp} (also references therein). It is relevant to note here that  there is no rigorous proof that the results found using the stochastic formalism will match with those found using quantum field  theory in the in-in formalism~\cite{Seery:2007wf}.   Via these computations, we may do hope to put some serious constraints on the fermion-inflaton Yukawa coupling.    \\

\section*{Acknowledgement} The authors would like to thank anonymous referee for making useful comments and suggestions.\\

\bigskip
\bigskip
\appendix

\labelformat{section}{Appendix #1} 
\section{Sketch of derivation of \ref{3p1}}
\label{appendix:a}

We wish to very briefly review in this appendix the derivation of the three point correlation function at equal time, \ref{3p1}, following~\cite{Motohashi:2012bb,Antoniadis:2011ib,Kehagias:2012pd}. We need to compute
\begin{equation}
    \xi^{(3)}_{f}(\vec{x}_{1},\vec{x}_{2},\vec{x}_3;t)=\langle f(\phi(\vec{x}_1,t))f(\phi(\vec{x}_2,t))f(\phi(\vec{x}_3,t))\rangle
\end{equation}
which equals
\begin{equation}
  \xi^{(3)}_{\phi}(\vec{x}_{1},\vec{x}_{2},\vec{x}_3;t)= \int  d\phi_1 d\phi_2 d\phi_3 \ f(\phi(\vec{x}_1,t))f(\phi(\vec{x}_2,t))f(\phi(\vec{x}_3,t))\rho_{3}(\phi_{1},\phi_{2},\phi_{3};t)  
\end{equation}
where $d\phi_{i}\equiv d\phi (\vec{x}_i,t)$, and  $\rho_3(\phi_{1},\phi_{2},\phi_{3};t)$ is the probability distribution function which satisfies a generalised version of the Fokker-Planck equation
\begin{equation}
\label{eq:fp-3}
    \frac{\partial\rho_3}{\partial t}=\sum_{i=1}^{3}\left[\frac{\partial}{\partial\phi_{i}}\left(\frac{V^{\prime}(\phi_{i})}{3H}\rho_{3}\right)+\frac{H^3}{8\pi^2}\frac{\partial^2\rho_3}{\partial\phi_i\partial\phi_i}\right]+\frac{H^3}{4\pi^2}\sum_{i,j; i\ensuremath{<} j}\frac{\partial^2\rho_3}{\partial\phi_i\partial\phi_j}j_{0}(\epsilon a(t)H|\vec{x}_i-\vec{x}_j|)
\end{equation}
For sub-Hubble modes, all the three points are well inside the Hubble radius ($H^{-1}$), and  we must have $a(t)|\vec{x}_i-\vec{x}_j|\ll H^{-1}$.  In that case we may set the Bessel function approximately to unity in \ref{eq:fp-3}. Then the equation permits a  static solution 
\begin{equation}
    \rho_3(\phi_{1},\phi_{2},\phi_{3};t)=\delta(\phi_3-\phi_2)\delta(\phi_2-\phi_1)\rho_{\text{eq}}(\phi_1)
    \label{A1}
\end{equation}
Since the initially sub-Hubble modes would eventually expand to super-Hubble scale, \ref{A1} can be treated as an initial condition for \ref{eq:fp-3}. With the help of  \ref{A1}, one finds after a little algebra the solution
\begin{equation}
  \xi^{(3)}_{f}(\vec{x}_{1},\vec{x}_{2},\vec{x}_3;t_1,t_2,t_3)=\mathcal{N}^2\sum_{a,b,c=1}^3 f_{a}f_{b}f_{c}B_{abc}
e^{-E_{a}(t-t_R)}e^{-E_{b}(t-t_r)}e^{-E_{c}(t-t_r)}e^{-E_{a}(t_r-t_R)},
\label{A2}
\end{equation}
where $B_{abc}$ and the  normalisation $\mathcal{N}$ is given by \ref{norm}, $f_{a}$ follows from the definition of spectral coefficients given in \ref{coefficient}.  $t_r$ and $t_R$ respectively denote the temporal coordinates, $t_r=-\ln{(\epsilon |\vec{x}_1-\vec{x}_2 | H)}/H$ and $t_R=-\ln{(\epsilon |\vec{x}_2-\vec{x}_3| H)}/H$. By defining next the proper separation, $R_{ij}=a(t)|\vec{x}_i-\vec{x}_j|$,  \ref{A2} can be recast as \ref{3p1}.
%

\section{Sketch of derivation of~\ref{bi-s3}}
\label{appendix:b}
We next wish to sketch the derivation of \ref{bi-s3}, starting from \ref{3p3}~\cite{Motohashi:2012bb,Antoniadis:2011ib,Kehagias:2012pd}.
Using 
\begin{eqnarray}
    \frac{1}{|\vec{x}|^{w_{i}}}=(2\pi)^{3}2^{2-w_{i}}\frac{\Gamma(\frac{3-w_{i}}{2})}{\Gamma(\frac{w_{i}}{2})}\int \frac{d^3\vec{k}}{(2\pi)^3}|\vec{k}|^{w_{i}-3}e^{i\vec{k}\cdot\vec{x}},
\end{eqnarray}
we have
\begin{equation}
\label{eq:ft-3}
    \xi_f^{(3)}(\vec{k}_1,\vec{k}_2,\vec{k}_3;a)=\prod_{m=i,j,k}\sum_{a,b,c}C_{abc}\ 2^{3-w_{m}}\pi^{3/2}\frac{\Gamma(\frac{3-w_{m}}{2})}{\Gamma(\frac{w_{m}}{2})}\frac{\delta^{(3)}(\vec{k}_1+\vec{k}_2+\vec{k}_3)}{(aH)^{w_{i}+w_{j}+w_{k}}}\int d^3\vec{k}|\vec{k}|^{w_k-3}|\vec{k}-\vec{k}_1|^{w_j-3}|\vec{k}+\vec{k}_2|^{w_i-3}, 
\end{equation}
which we rewrite as
\begin{equation}\label{eq:3pt-bs}
\xi_f^{(3)}(\vec{k}_1,\vec{k}_2,\vec{k}_3;a)=(2\pi)^3\delta^{(3)}(\vec{k}_1+\vec{k}_2+\vec{k}_3)B_{S}(\vec{k}_{1},\vec{k}_{2}, \vec{k}_{3};a),
\end{equation}
where $B_{S}(\vec{k}_{1}, \vec{k}_{2}, \vec{k}_{3};a)$, known as  the scalar bispectrum, explicitly reads 
\begin{equation}\label{bi-s1}
\begin{split}
B_{S}(\vec{k}_{1}, \vec{k}_{2}, \vec{k}_{3};a)&=\pi^{5/2}\sum_{a,b,c} C_{abc}\frac{2^{7-(w_i+w_j+w_k)}}{(aH)^{w_{i}+w_{j}+w_{k}}}\frac{\Gamma(3-\frac{w_{i}+w_{j}+w_{k}}{2})\Gamma(\frac{3-w_k}{2})}{\Gamma(\frac{w_i}{2})\Gamma(\frac{w_j}{2})}\\
&\times k_{1}^{w_{i}+w_{j}+w_{k}-6}\int_{0}^{1}ds\frac{(1-s)^{\frac{1}{2}-\frac{w_i}{2}}s^{\frac{1}{2}-\frac{w_j}{2}}}{{\left[(1-s)X+sY\right]}^{3-\frac{w_{i}+w_{j}+w_{k}}{2}}} {}\ _2F_{1}\left( 3-\frac{w_{i}+w_{j}+w_{k}}{2},\frac{w_{k}}{2},\frac{3}{2},{Z}\right)
\end{split}
\end{equation}
where $X$, $Y$ and $Z$ are given by
\begin{eqnarray}
\label{XYZ}
    X=\frac{k_{2}^{2}}{k_{1}^{2}},{\hspace{0.5cm}} Y=\frac{k_{3}^{2}}{k_{1}^{2}}, {\hspace{0.5cm}} {Z}=1-\frac{s(1-s)}{(1-s)X+sY}
\end{eqnarray}
Using now the transformation formula, 
\begin{equation*}
    {}_{2}F_{1}\left(a,b,c,x\right)=(1-x)^{c-a-b}{}_2F_{1}\left(c-a,c-b,c,x\right),
\end{equation*}
\ref{bi-s1} can be recast as \ref{bi-s3}.

\bigskip


\begin{thebibliography}{99} 

%
\bibitem{Mukhanov:2005sc}
V.~Mukhanov,
{\it Physical Foundations of Cosmology},
Cambridge University Press, (2005).
%
%
\bibitem{Starobinsky:1982ee}
A.~A.~Starobinsky,
{\it Dynamics of Phase Transition in the New Inflationary Universe Scenario and Generation of Perturbations},
Phys. Lett. B\textbf{117}, 175 (1982).
%
%
\bibitem{Tsamis}
N. C. Tsamis and R. P. Woodard, 
{\it Relaxing the cosmological constant}, Phys. Lett. B{\bf301}, 351 (1993).
%
%
\bibitem{Ringeval}
C. Ringeval, T. Suyama, T. Takahashi, M. Yamaguchi and S. Yokoyama, 
{\it Dark energy from primordial inflationary quantum 
fluctuations},
Phys. Rev. Lett.{\bf105}, 121301 (2010) [arXiv:1006.0368 [astro-ph.CO]].
%
%
\bibitem{Evnin:2018zeo}
O.~Evnin and K.~Nguyen,
{\it Graceful exit for the cosmological constant damping scenario},
Phys. Rev. D\textbf{98}, no.12, 124031 (2018)
[arXiv:1810.12336 [gr-qc]].
%
%
\bibitem{Allen:1985ux}
B.~Allen,
{\it Vacuum States in de Sitter Space},
Phys. Rev. D\textbf{32}, 3136 (1985).
%
%
\bibitem{Allen}
B. Allen and A. Folacci,
{\it Massless minimally coupled scalar field in de Sitter space},
Phys. Rev. D\textbf{35}, 3771 (1987).
%
%
\bibitem{Floratos}
E. G. Floratos, J. Iliopoulos and T. N. Tomaras, 
{\it Tree Level Scattering Amplitudes in De Sitter Space diverge}, Phys. Lett. B{\bf197}, 373 (1987).
%
%
\bibitem{Onemli:2002hr}
V.~K.~Onemli and R.~P.~Woodard,
{\it Superacceleration from massless, minimally coupled $\phi^4$},
Class.~Quant.~Grav.\textbf{19}, 4607 (2002)
[arXiv:gr-qc/0204065 [gr-qc]].
%
%
\bibitem{Weinberg}
S. Weinberg, 
{\it Quantum contributions to cosmological correlations}, Phys. Rev. D\textbf{72}, 043514 (2005).
%
%
\bibitem{Brunier:2004sb} 
  T.~Brunier, V.~K.~Onemli and R.~P.~Woodard,
{\it Two loop scalar self-mass during inflation}.
  Class.\ Quant.\ Grav.{\bf 22}, 59 (2005)
  [gr-qc/0408080].
%
%
\bibitem{Miao:2006pn}
  S.~P.~Miao and R.~P.~Woodard,
{\it Leading log solution for inflationary Yukawa},
  Phys.\ Rev.\ D{\bf 74}, 044019 (2006)
  [gr-qc/0602110].
 %
 %
 \bibitem{Garbrecht:2006jm}
B.~Garbrecht and T.~Prokopec,
{\it Fermion mass generation in de Sitter space},
Phys. Rev. D\textbf{73}, 064036 (2006)
[arXiv:gr-qc/0602011 [gr-qc]].
 %
 %
 \bibitem{Enqvist:2008kt}
K.~Enqvist, S.~Nurmi, D.~Podolsky and G.~I.~Rigopoulos,
{\it On the divergences of inflationary superhorizon perturbations},
JCAP\textbf{04}, 025 (2008)
[arXiv:0802.0395 [astro-ph]].
%
 %
\bibitem{Onemli:2015pma}
V.~K.~Onemli,
{\it Vacuum Fluctuations of a Scalar Field during Inflation: Quantum versus Stochastic Analysis},
Phys.~Rev.~D\textbf{91}, 103537 (2015)
[arXiv:1501.05852 [gr-qc]].
%
%
\bibitem{Burgess:2015ajz}
C.~P.~Burgess, R.~Holman and G.~Tasinato,
{\it Open EFTs, IR effects $\&$ late-time resummations: systematic corrections in stochastic inflation},
JHEP\textbf{01}, 153 (2016)
[arXiv:1512.00169 [gr-qc]].
%
%
\bibitem{Karakaya:2017evp}
G.~Karakaya and V.~K.~Onemli,
{\it Quantum effects of mass on scalar field correlations, power spectrum, and fluctuations during inflation},
Phys.~Rev.~D\textbf{97}, no.12, 123531 (2018)
[arXiv:1710.06768 [gr-qc]].
%
%
\bibitem{Kamenshchik:2020yyn}
A.~Y.~Kamenshchik and T.~Vardanyan,
{\it Renormalization group inspired autonomous equations for secular effects in de Sitter space},
Phys. Rev. D\textbf{102}, no.6, 065010 (2020)
[arXiv:2005.02504 [hep-th]].
%
\bibitem{Kamenshchik:2024ybm}
A.~Kamenshchik and P.~Petriakova,
{\it IR finite correlation functions in de Sitter space, a smooth massless limit, and an autonomous equation},
[arXiv:2410.16226 [hep-th]].
%
%
\bibitem{Miao:2021gic}
S.~P.~Miao, N.~C.~Tsamis and R.~P.~Woodard,
{\it Summing inflationary logarithms in nonlinear sigma models},
JHEP\textbf{03}, 069 (2022)
[arXiv:2110.08715 [gr-qc]].
%
%
\bibitem{Bhattacharya:2022aqi}
S.~Bhattacharya,
{\it Massless minimal quantum scalar field with an asymmetric self interaction in de Sitter spacetime},
JCAP\textbf{09}, 041 (2022)
[arXiv:2202.01593 [hep-th]].
%
%
\bibitem{Bhattacharya:2023yhx}
S.~Bhattacharya and M.~D.~Choudhury,
{\it Non-perturbative $\langle \phi \rangle$, $\langle \phi^2 \rangle$ and the dynamically generated scalar mass with Yukawa interaction in the inflationary de Sitter spacetime},
JCAP\textbf{01}, 056 (2024)
[arXiv:2308.11384 [hep-th]].
%
%
\bibitem{Cabass:2022avo}
G.~Cabass, M.~M.~Ivanov, M.~Lewandowski, M.~Mirbabayi and M.~Simonovi\'c,
    {\it Snowmass white paper: Effective field theories in cosmology},
Phys. Dark Univ.\textbf{40}, 101193 (2023)
[arXiv:2203.08232 [astro-ph.CO]].
%
\bibitem{Brahma:2024yor}
S.~Brahma, J.~Calder\'on-Figueroa and X.~Luo,
{\it Time-convolutionless cosmological master equations: Late-time resummations and decoherence for non-local kernels},
[arXiv:2407.12091 [hep-th]].
%
\bibitem{Akhmedov:2024npw}
E.~T.~Akhmedov, V.~I.~Lapushkin and D.~I.~Sadekov,
{\it Light fields in various patches of de Sitter space-time},
[arXiv:2411.11106 [hep-th]].
%


%
\bibitem{Starobinsky:1986fx}
A.~A.~Starobinsky,
{\it Stochastic  de Sitter (inflationary) stage in the early universe},
Lect. Notes Phys.\textbf{246}, 107 (1986).
%
%
\bibitem{Starobinsky:1994bd}
A.~A.~Starobinsky and J.~Yokoyama,
{\it Equilibrium state of a selfinteracting scalar field in the De Sitter background},
Phys. Rev. D\textbf{50}, 6357 (1994)
[arXiv:astro-ph/9407016 [astro-ph]].
%
%
\bibitem{Finelli:2008zg}
F.~Finelli, G.~Marozzi, A.~A.~Starobinsky, G.~P.~Vacca and G.~Venturi,
{\it Generation of fluctuations during inflation: Comparison of stochastic and field-theoretic approaches},
Phys.~Rev.~D\textbf{79}, 044007 (2009)
[arXiv:0808.1786 [hep-th]].
%
%
\bibitem{Martin:2011ib}
J.~Martin and V.~Vennin,
{\it Stochastic Effects in Hybrid Inflation},
Phys. Rev. D\textbf{85}, 043525 (2012)
[arXiv:1110.2070 [astro-ph.CO]].
%
%
\bibitem{Motohashi:2012bb}
H.~Motohashi, T.~Suyama and J.~Yokoyama,
{\it Consequences of a stochastic approach to the conformal invariance of inflationary correlators},
Phys. Rev. D\textbf{86}, 123514 (2012)
[arXiv:1210.2497 [hep-th]].
%


%
\bibitem{Fujita:2013cna}
T.~Fujita, M.~Kawasaki, Y.~Tada and T.~Takesako,
{\it  A new algorithm for calculating the curvature perturbations in stochastic inflation},
JCAP\textbf{12}, 036 (2013)
[arXiv:1308.4754 [astro-ph.CO]].
%
%
\bibitem{Fujita:2014tja}
T.~Fujita, M.~Kawasaki and Y.~Tada,
{\it Non-perturbative approach for curvature perturbations in stochastic $\delta N$ formalism},
JCAP\textbf{10}, 030 (2014)
[arXiv:1405.2187 [astro-ph.CO]].
%
%
\bibitem{Vennin:2015hra}
V.~Vennin and A.~A.~Starobinsky,
{\it Correlation Functions in Stochastic Inflation},
Eur. Phys. J. C\textbf{75}, 413 (2015)
[arXiv:1506.04732 [hep-th]].
%
%
\bibitem{Assadullahi:2016gkk}
H.~Assadullahi, H.~Firouzjahi, M.~Noorbala, V.~Vennin and D.~Wands,
{\it Multiple Fields in Stochastic Inflation},
JCAP\textbf{06}, 043 (2016)
[arXiv:1604.04502 [hep-th]].
%
%
\bibitem{Firouzjahi:2018vet}
H.~Firouzjahi, A.~Nassiri-Rad and M.~Noorbala,
{\it Stochastic Ultra Slow Roll Inflation},''
JCAP\textbf{01}, 040 (2019)
[arXiv:1811.02175 [hep-th]].
%
%
\bibitem{Pattison:2019hef}
C.~Pattison, V.~Vennin, H.~Assadullahi and D.~Wands,
{\it Stochastic inflation beyond slow roll},
JCAP\textbf{07}, 031 (2019)
[arXiv:1905.06300 [astro-ph.CO]].
%

%
\bibitem{Hardwick:2019uex}
R.~J.~Hardwick, T.~Markkanen and S.~Nurmi,
{\it Renormalisation group improvement in the stochastic formalism},
JCAP\textbf{09}, 023 (2019)
%
%
\bibitem{Markkanen:2019kpv}
T.~Markkanen, A.~Rajantie, S.~Stopyra and T.~Tenkanen,
{\it Scalar correlation functions in de Sitter space from the stochastic spectral expansion},
JCAP\textbf{08}, 001 (2019)
[arXiv:1904.11917 [gr-qc]].
%
%
\bibitem{Markkanen:2020bfc}
T.~Markkanen and A.~Rajantie,
{\it Scalar correlation functions for a double-well potential in de Sitter space},
JCAP\textbf{03}, 049 (2020)
[arXiv:2001.04494 [gr-qc]].
%
%
\bibitem{Ando:2020fjm}
K.~Ando and V.~Vennin,
{\it Power spectrum in stochastic inflation},
JCAP\textbf{04}, 057 (2021)
[arXiv:2012.02031 [astro-ph.CO]].
%
%
\bibitem{Ebadi:2023xhq}
R.~Ebadi, S.~Kumar, A.~McCune, H.~Tai and L.~T.~Wang,
{\it Gravitational Waves from Stochastic Scalar Fluctuations},
[arXiv:2307.01248 [astro-ph.CO]].
%
%
\bibitem{Cruces:2022imf}
D.~Cruces,
{\it Review on Stochastic Approach to Inflation},
Universe\textbf{8}, no.6, 334 (2022)
[arXiv:2203.13852 [gr-qc]].
%

%
\bibitem{Gangui:1993tt}
A.~Gangui, F.~Lucchin, S.~Matarrese and S.~Mollerach,
{\it The Three point correlation function of the cosmic microwave background in inflationary models},
Astrophys. J.\textbf{430}, 447 (1994)
[arXiv:astro-ph/9312033 [astro-ph]].
%
%
\bibitem{Verde:1999ij}
L.~Verde, L.~M.~Wang, A.~Heavens and M.~Kamionkowski,
{\it Large scale structure, the cosmic microwave background, and primordial non-gaussianity},
Mon. Not. Roy. Astron. Soc.\textbf{313}, L141(2000)
[arXiv:astro-ph/9906301 [astro-ph]].
%
%
\bibitem{Komatsu:2001rj}
E.~Komatsu and D.~N.~Spergel,
{\it Acoustic signatures in the primary microwave background bispectrum},
Phys. Rev. D\textbf{63}, 063002 (2001)
[arXiv:astro-ph/0005036 [astro-ph].
%
%
\bibitem{Maldacena:2002vr}
J.~M.~Maldacena,
{\it Non-Gaussian features of primordial fluctuations in single field inflationary models},
JHEP\textbf{05}, 013 (2003)
[arXiv:astro-ph/0210603 [astro-ph]].
%
%
\bibitem{Creminelli:2004yq}
P.~Creminelli and M.~Zaldarriaga,
{\it Single field consistency relation for the 3-point function},
JCAP\textbf{10}, 006 (2004)
[arXiv:astro-ph/0407059 [astro-ph]].
%
%
\bibitem{Lyth:2005fi}
D.~H.~Lyth and Y.~Rodriguez,
{\it The Inflationary prediction for primordial non-Gaussianity},
Phys. Rev. Lett.\textbf{95}, 121302 (2005)
[arXiv:astro-ph/0504045 [astro-ph]].
%
%
\bibitem{Boubekeur:2005fj}
L.~Boubekeur and D.~H.~Lyth,
{\it Detecting a small perturbation through its non-Gaussianity},
Phys.~Rev.~D\textbf{73}, 021301 (2006)
[arXiv:astro-ph/0504046 [astro-ph]].
%
%
\bibitem{Meerburg:2009ys}
P.~D.~Meerburg, J.~P.~van der Schaar and P.~S.~Corasaniti,
{\it Signatures of Initial State Modifications on Bispectrum Statistics},
JCAP\textbf{05}, 018 (2009)
[arXiv:0901.4044 [hep-th]].
%
%
\bibitem{Jeong:2009vd}
D.~Jeong and E.~Komatsu,
{\it Primordial non-Gaussianity, scale-dependent bias, and the bispectrum of galaxies},
Astrophys. J.\textbf{703}, 1230-1248 (2009)
[arXiv:0904.0497 [astro-ph.CO]].
%
%
\bibitem{yadav2010primordial}
A.~P.~S.~Yadav and  B.~D.~Wandelt, 
{\it Primordial Non-Gaussianity in the Cosmic Microwave Background}, {Advances in Astronomy}\textbf{1}, p565248
 (2010).
%
%
\bibitem{Antoniadis:2011ib}
I.~Antoniadis, P.~O.~Mazur and E.~Mottola,
{\it Conformal Invariance, Dark Energy, and CMB Non-Gaussianity},
JCAP\textbf{09}, 024 (2012)
[arXiv:1103.4164 [gr-qc]].
%
%
\bibitem{Kehagias:2012pd}
A.~Kehagias and A.~Riotto,
{\it Operator Product Expansion of Inflationary Correlators and Conformal Symmetry of de Sitter},
Nucl. Phys. B\textbf{864}, 492-529 (2012)
[arXiv:1205.1523 [hep-th]].
%
%
\bibitem{Gwyn:2012pb}
R.~Gwyn, M.~Rummel and A.~Westphal,
{\it Resonant non-Gaussianity with equilateral properties},
JCAP\textbf{04}, 040 (2013)
[arXiv:1211.0070 [hep-th]].
%
%
\bibitem{Kristiano:2023scm}
J.~Kristiano and J.~Yokoyama,
{\it Note on the bispectrum and one-loop corrections in single-field inflation with primordial black hole formation},
Phys. Rev. D\textbf{109}, no.10, 103541 (2024)
[arXiv:2303.00341 [hep-th]].
%




\bibitem{Enqvist:2012xn}
K.~Enqvist, R.~N.~Lerner, O.~Taanila and A.~Tranberg,
{\it Spectator field dynamics in de Sitter and curvaton initial conditions},
JCAP\textbf{10}, 052 (2012)
[arXiv:1205.5446 [astro-ph.CO]].
%
%
\bibitem{Friedrich:2019hev}
P.~Friedrich and T.~Prokopec,
{\it Entropy production in inflation from spectator loops},
Phys. Rev. D\textbf{100}, no.8, 083505 (2019)
[arXiv:1907.13564 [astro-ph.CO]].
%
%
\bibitem{Rigopoulos:2022gso}
G.~Rigopoulos and A.~Wilkins,
{\it Computing first-passage times with the functional renormalisation group},
JCAP\textbf{04}, 046 (2023)
[arXiv:2211.09649 [astro-ph.CO]].
%
%
\bibitem{Glavan:2023lvw}
D.~Glavan and T.~Prokopec,
{\it When tadpoles matter: one-loop corrections for spectator Higgs in inflation},
JHEP\textbf{10}, 063 (2023)
[arXiv:2306.11162 [hep-ph]].
%
%
\bibitem{Bhattacharya:2023twz}
S.~Bhattacharya and N.~Joshi,
{\it Decoherence and entropy generation at one loop in the inflationary de Sitter spacetime for Yukawa interaction},
JCAP\textbf{04}, 078 (2024)
[arXiv:2307.13443 [hep-th]].
%
%
\bibitem{Wilkins:2023asp}
A.~Wilkins and A.~Cable,
{\it Spectators no more! How even unimportant fields can ruin your Primordial Black Hole model},
JCAP\textbf{02}, 026 (2024)
[arXiv:2306.09232 [astro-ph.CO]].
%
%
\bibitem{Palma:2023idj}Belfiglio:2024xqt
G.~A.~Palma and S.~Sypsas,
{\it Non-Gaussian statistics of de Sitter spectators: A perturbative derivation of stochastic dynamics},
[arXiv:2309.16474 [hep-th]].
%
%
\bibitem{Belfiglio:2024xqt}
A.~Belfiglio and O.~Luongo,
{\it Production of ultralight dark matter from inflationary spectator fields},
Phys. Rev. D\textbf{110}, no.2, 023541 (2024)
[arXiv:2401.16910 [hep-th]].
%

%
  \bibitem{Candelas}
  P.~Candelas and D.~J.~Raine, 
  {\it General-relativistic quantum field theory: An exactly soluble model}, Phys.~Rev.~D{\bf 12}  965 (1975).
%

%
\bibitem{Planck:2018jri}
Y.~Akrami \textit{et al.} [Planck],
{\it Planck 2018 results. X. Constraints on inflation},
Astron. Astrophys.\textbf{641}, A10 (2020)
[arXiv:1807.06211 [astro-ph.CO]].
%
%
\bibitem{Planck:2019kim}
Y.~Akrami \textit{et al.} [Planck],
{\it Planck 2018 results. IX. Constraints on primordial non-Gaussianity},
Astron. Astrophys.\textbf{641}, A9 (2020)
[arXiv:1905.05697 [astro-ph.CO]].
%
%
\bibitem{Koksma:2009tc}
J.~F.~Koksma and T.~Prokopec,
{\it Fermion Propagator in Cosmological Spaces with Constant Deceleration},
Class. Quant. Grav. \textbf{26}, 125003 (2009)
[arXiv:0901.4674 [gr-qc]].


\bibitem{Abolhasani:2011yp}
A.~A.~Abolhasani, H.~Firouzjahi and M.~Sasaki,
{\it Curvature perturbation and waterfall dynamics in hybrid inflation}
JCAP \textbf{10}, 015 (2011)
[arXiv:1106.6315 [astro-ph.CO]].

\bibitem{Seery:2007wf}
D.~Seery,
{\it One-loop corrections to the curvature perturbation from inflation},
JCAP \textbf{02}, 006 (2008)
[arXiv:0707.3378 [astro-ph]].

\end{thebibliography}
\end{document}